\documentclass[useAMS,usenatbib]{mn2e}
\usepackage{amssymb,epsfig,natbib}
\title[A deeper search for the progenitor of the Type Ic Supernova 2002ap]
{A deeper search for the progenitor of the Type Ic Supernova 2002ap}
\author[R. M. Crockett {\rm et al.}]
{R. M. Crockett$^1$ \thanks{E-mail: rcrockett02@qub.ac.uk}, S. J. Smartt$^1$, J. J. Eldridge$^1$, S. Mattila$^1$, D. R. Young$^1$,
\newauthor A. Pastorello$^1$, J. R. Maund$^2$, C. R. Benn$^3$, I. Skillen$^3$\\ 
\\
$^1$Astrophysics Research Centre, School of Maths and Physics, Queen's University Belfast, 
BT7 1NN, Northern Ireland, UK\\
$^2$Department of Astronomy and McDonald Observatory, University of Texas, 1 University Station, C1400, Austin, TX, 78712, U.S.A.\\
$^3$Isaac Newton Group, Apartado 321, E-38700 Santa Cruz de La Palma, Spain\\
}
\date{Submitted 04 June 2007}

\pubyear{2007}

\def\LaTeX{L\kern-.36em\raise.3ex\hbox{a}\kern-.15em
  T\kern-.1667em\lower.7ex\hbox{E}\kern-.125emX}

\def \aj {AJ}
\def \mnras {MNRAS}
\def \apj {ApJ}
\def \apjl {ApJL}
\def \aap {A\&A}
\def \nat {Nature}
\def \araa {ARAA}
\def \iauc {IAUC}
\def \pasp {PASP}

\def \apjs {ApJS}
\def \aaps {A\&AS}

\bibpunct{(}{)}{;}{a}{}{,}

\newcommand{\ebv}{\mbox{$E(B\!-\!V)$}}

\begin{document}

\label{firstpage}

\maketitle
\begin{abstract}

Images of the site of the Type Ic Supernova 2002ap taken before explosion were analysed previously by Smartt et al. (2002). We have uncovered new unpublished, archival pre-explosion images from the Canada-France-Hawaii Telescope (CFHT) that are vastly superior in depth and image quality. In this paper we present a further search for the progenitor star of this unusual Type Ic supernova. Aligning high-resolution Hubble Space Telescope (HST) observations of the supernova itself with the archival CFHT images allowed us to pinpoint the location of the progenitor site on the ground based observations. We find that a source visible in the $B$ and $R$ band pre-explosion images close to the position of the SN is (1) not coincident with the SN position within the uncertainties of our relative astrometry, and (2) is still visible $\sim$4.7 yrs post explosion in late-time observations taken with the William Herschel Telescope. We therefore conclude that it is not the progenitor of SN~2002ap. We derived absolute limiting magnitudes for the progenitor of \mbox{$M_B$ $\geq$ -4.2 $\pm$ 0.5} and \mbox{$M_R$ $\geq$ -5.1 $\pm$ 0.5}. These are the deepest limits yet placed on a Type Ic supernova progenitor. We rule out all massive stars with initial masses greater than 7-8$M_{\odot}$ (the lower mass limit for stars to undergo core collapse) that have not evolved to become Wolf-Rayet stars. This is consistent with the prediction that Type Ic supernovae should result from the explosions of Wolf-Rayet stars. Comparing our luminosity limits with stellar models of single stars at appropriate metallicity (Z=0.008) and with {\em standard} mass loss rates, we find no model that produces a Wolf-Rayet star of low enough mass and luminosity to be classed as a viable progenitor. Models with twice the standard mass loss rates provide possible single star progenitors but all are initially more massive than 30-40$M_{\odot}$. We conclude that any single star progenitor must have experienced at least twice the standard mass loss rates, been initially more massive than $30-40M_{\odot}$ and exploded as a W-R star of final mass 10-12$M_{\odot}$. Alternatively a progenitor star of lower initial mass may have evolved in an interacting binary system. Mazzali et al. (2002) propose such a binary scenario for the progenitor of SN~2002ap in which a star of initial mass 15-20$M_{\odot}$ is stripped by its binary companion, becoming a 5$M_{\odot}$ Wolf-Rayet star prior to explosion. We constrain any possible binary companion to a main sequence star of $\leq20M_{\odot}$, a neutron star or a black hole. By combining the pre-explosion limits with the ejecta mass estimates and constraints from X-ray and radio observations we conclude that any binary interaction most likely occurred as Case B mass transfer, either with or without a subsequent common envelope evolution phase.

\end{abstract}


\section{Introduction}

Stellar evolution theory predicts that stars initially more massive than 7-8$M_{\odot}$ should explode as Core Collapse Supernovae (CCSNe) at the end of their nuclear burning lives (e.g. Heger et al. 2003; Eldridge \& Tout 2004). The observation of progenitor stars of CCSNe in archival pre-explosion images, provides us with a direct means with which to test the theoretical predictions. Until quite recently the only directly observed progenitor stars of CCSNe were those of the Type II-peculiar SN~1987A in the Large Magellanic Cloud (LMC) (Walborn et al. 1987) and the Type IIb SN~1993J in M~81 (Aldering et al 1994). The advent of large data archives, perhaps most importantly that of the Hubble Space Telescope (HST), has allowed several groups around the world to extend the progenitor search to larger distances (e.g. Smartt et al. 2004; Li et al. 2005; Gal-Yam et al. 2007). To date a total of nine progenitor stars of CCSNe have been reported, all of which are for hydrogen-rich Type II SNe, and with the seven most recent additions specifically for Type II-P SNe. There has been as yet no direct detection of a hydrogen-deficient Type Ib or a hydrogen and helium deficient Type Ic supernova progenitor, although the precursor of the peculiar SN~2006jc was observed some two years prior to explosion during an LBV-like outburst (Pastorello et al. 2007). Upper limits have been determined for the luminosity of several Type Ib/c progenitors, of which the most restrictive limits set thus far are for the progenitor of the Type Ic SN~2004gt (Maund, Smartt \& Schweizer 2005; Gal-Yam et al. 2005). In terms of absolute magnitude, the pre-explosion HST observations probed down to $M_V$ = -5.3. This allowed both groups to determine that the progenitor was almost certainly a Wolf-Rayet star, a massive star that has lost its hydrogen-rich envelope. Such stars are predicted to be the progenitors of Type Ib/c SNe.

Another Type Ic SN with archival pre-explosion observations is SN~2002ap. It was discovered by Y. Hirose on 2002 January 29.4 UT at magnitude V=14.54 in the spiral galaxy M~74 (Nakano et al. 2002), at a distance of approximately 9.3 Mpc (Hendry et al. 2005). The object was quickly revealed as a Type Ic SN with broad spectral features, indicative of very high velocities of the SN ejecta (e.g. Meikle et al. 2002; Kinugasa et al. 2002; Gal-Yam et al. 2002; Mazzali et al. 2002; Foley et al. 2003). Although its spectra appeared similar to that of the peculiar SN 1998bw, it was less luminous, reaching a peak magnitude of $M_V\simeq$ -17.5, about 1.7 mag fainter than SN~1998bw. Furthermore, unlike SN~1998bw no gamma ray burst (GRB) was detected coincident with the position of SN~2002ap (Hurley et al. 2002a). $UBVRIKH\alpha$ pre-explosion observations of the site of SN~2002ap in M~74, taken with the KPNO 0.9~m, the 2.5~m Isaac Newton Telescope (INT) and the Bok 2.3~m telescope, were used by Smartt et al. (2002) to derive upper limits to the absolute brightness of the progenitor star in the various filters. The $B$ and $V$ band observations were the deepest, reaching absolute magnitude limits as faint as $M_B$ = -6.3 and $M_V$ = -6.6. 

At the time it was believed that these were the best quality images available of the pre-explosion site of SN~2002ap. However we have discovered additional 
pre-explosion images of the SN position in the archive of the Canada France Hawaii Telescope (CFHT), which are vastly superior in depth and resolution to those originally presented in Smartt et al. (2002). In fact we shall show that they are the deepest pre-explosion images of any nearby Type Ib/c SN to date. 
The new archive at the Canadian Astronomy Data Centre (CADC)\footnote{http://cadcwww.dao.nrc.ca/cfht/} returns these images. This paper presents the study of these very deep, high quality ground based observations of the pre-explosion site of SN~2002ap, in which we attempt to detect the progenitor star and place rigorous constraints on its properties.

\section{Observations}\label{sec:obs}

\subsection{Pre-explosion observations}\label{sec:pre}

The galaxy M74 was observed with the 3.6-m Canada France Hawaii Telescope in October 1999, and to our knowledge the observations have not been published. The site of SN~2002ap was imaged using CFH12K, a 12 CCD mosaic camera, which is now decommissioned. The pixel size of this instrument was 0\farcs2 and further details can be found on the CFHT website\footnote{http://www.cfht.hawaii.edu/Instruments/Imaging/CFH12K/}. The $B$ and $R$ observations were particularly deep, consisting of individual, dithered 600-second exposures which combined to result in 1 hour 20 mins and 1 hour exposure times respectively. The $V$ and $H\alpha$ observations were much shallower with the individual frames combined giving total exposure times of just 600 and 900-seconds respectively. A summary of these observations can be found in Table \ref{tab:obs}. The images were downloaded from the CADC archive, and most had been processed successfully through the ELIXIR system. However, some of the $R$ band images generated by ELIXIR were corrupted, so we used the master flat from the archive and processed the frames manually using standard techniques in IRAF. 

There were 8$\times$600s exposures from 1999 October 9th in $B$ and the combined image quality was 0\farcs75. A further 8$\times$600s set of exposures were taken on 1999 December 31st, but seven of these had an image quality of $\sim$1\farcs2 (the remaining image had poor seeing of 1\farcs9). Combining the 1\farcs2 stacked image with the better seeing 0\farcs75 images did not increase the sensitivity, so we chose to proceed with analysis of the 1999 October 9th images only. The 6$\times$600s $R$ band images taken on 1999 October 6th were all of similar resolution (0\farcs7), and the shorter 5$\times$120s $V$ stack had a resultant quality of 1\farcs2. There were 5$\times$300s $H\alpha$ exposures taken on 1999 Jul 13, each with image quality around 0\farcs9. The observations were made towards the end of the night which resulted in the final two exposures having significantly higher background counts. These two exposures were excluded from the final stacked image since they would only serve to add excess noise.  

\begin{figure}
  \begin{center}
    \epsfig{file = 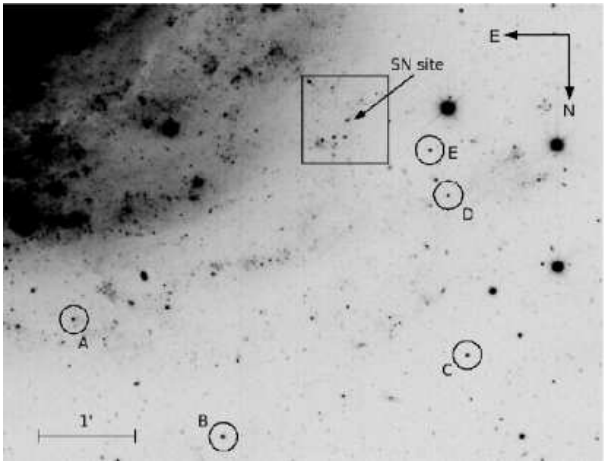, width = 80mm}
    \caption{Standard stars in CFHT pre-explosion observations. The approximate location of the site of SN~2002ap is also indicated. All other images in this paper have the same orienation as this figure.}
  \label{fig:finder}
  \end{center}
\end{figure}

To determine zeropoints and colour corrections for each filter, magnitudes of standard stars identified from the catalogue of Henden et al. (2002) were measured using the aperture photometry task within the {\sc iraf} {\sc daophot} package. Due to the depth of the observations many of the standard stars in the field were saturated and it was only possible to find five useable objects common in {\it BVR}. These stars are shown labelled in Figure~\ref{fig:finder} and the corresponding catalogue magnitudes are presented in Table \ref{tab:standards}. The calculated zeropoints are shown in Table \ref{tab:zero}.

\begin{table*}
  \caption[]{Summary of observational data.}
   \begin{center}
      \begin{tabular}{lrrrr} \hline\hline
        Date of Observation & Telescope + & Filter & Exposure Time & Observer\\
	& Instrument & & (sec)\\
        \hline
Pre-explosion observations\\[1.5ex]
1999 Jul 13 & CFHT+CFH12K & H$\alpha$ & 900 & Cuillandre, Hainaut, McDonald\\
1999 Oct 06 & CFHT+CFH12K & R & 3600 & Astier, Fabbro, Pain\\
1999 Oct 09 & CFHT+CFH12K & B & 4800 & McDonald, Cuillandre, Veillet\\
1999 Oct 09 & CFHT+CFH12K & V & 600 & McDonald, Cuillandre, Veillet\\[1.5ex]
	\hline
Post-explosion observations\\[1.5ex]
2003 Jan 10 & HST+ACS/HRC & F475W & 120 & Kirshner\\
2003 Jan 10 & HST+ACS/HRC & F625W & 120 & Kirshner\\
2003 Jan 10 & HST+ACS/HRC & F658N & 800 & Kirshner\\
2003 Jan 10 & HST+ACS/HRC & F814W & 180 & Kirshner\\
2004 Jul 06 & HST+ACS/HRC & F555W & 480 & Filippenko\\
2004 Jul 06 & HST+ACS/HRC & F814W & 720 & Filippenko\\
2004 Aug 31 & HST+ACS/HRC & F435W & 840 & Filippenko\\
2004 Aug 31 & HST+ACS/HRC & F625W & 360 & Filippenko\\
2006 Aug 25 & WHT+AUX & B & 4800 & Benn, Skillen\\
2006 Aug 27 & WHT+AUX & R & 3600 & Benn, Skillen\\
        \hline\hline
      \end{tabular}\\
    \end{center}
    {\footnotesize CFHT = 3.6-m Canada France Hawaii Telescope, Mauna Kea, Hawaii\\
HST+ACS/HRC = Hubble Space Telescope + Advanced Camera for Surveys/High Resolution Camera\\
WHT = 4.2-m William Herschel Telescope, La Palma, Canary Islands}
  \label{tab:obs}
\end{table*}

\begin{table}
  \caption[]{Standard star magnitudes taken from Henden et al. (2002).}
    \begin{center}
      \begin{tabular}{rcccc} \hline\hline
        Star & Henden catalogue & B & V & R\\
	& identity\\
        \hline
A & 118 & 20.827 & 19.286 & 18.257\\ 
B & 105 & 20.231 & 19.756 & 19.523\\
C & 91 & 20.804 & 19.286 & 18.234\\ 
D & 87 & 21.333 & 19.827 & 18.696\\ 
E & 85 & 20.739 & 20.536 & 20.326\\ 
        \hline\hline
      \end{tabular}\\
    \end{center}
  \label{tab:standards}
\end{table}

\begin{table}
  \caption[]{CFHT zeropoints and colour corrections.}
    \begin{center}
      \begin{tabular}{lrr} \hline\hline
        Filter & Colour Correction & Zeropoint\\
	& (Colour)\\
        \hline
B & 0.04(B-R) & 26.10\\ 
R & 0.02(B-R) & 26.29\\
        \hline\hline
      \end{tabular}\\
    \end{center}
  \label{tab:zero}
\end{table}

\subsection{Post-explosion observations}\label{sec:post}

Observations of SN~2002ap were taken using the High Resolution Channel (HRC) (pixel scale = 0\farcs025) of the Advanced Camera for Surveys (ACS) on board HST during January 2003, and July and August 2004. These formed part of the programmes GO9114 (PI: R. Kirshner) and SNAP10272 (PI: A. Filippenko), designed to obtain late-time photometry of nearby SNe. Details of these observations are given in Table \ref{tab:obs}. All HST observations were downloaded from the Space Telescope Science Institute (STScI) archive via the on-the-fly re-calibration (OTFR) pipeline. Those from 2004 were found to be well aligned despite being taken on two different epochs almost two months apart. Identification of SN~2002ap in these late-time images was aided by comparison with those from January 2003 in which the SN was significantly brighter.  In all cases the pointing of the observations was such that SN~2002ap was imaged very close to the centre of the HRC chip.  

Photometry was carried out using the ACS module of the PSF-fitting photometry package {\sc dolphot}\footnote{http://purcell.as.arizona.edu/dolphot/}, a modified version of HSTphot (Dolphin 2000). This package uses model PSFs to automatically detect, characterise and perform photometry of objects in ACS images. Furthermore it incorporates aperture corrections, charge transfer efficiency (CTE) corrections (Reiss 2003) and transformation of HST flight system magnitudes to the standard Johnson-Cousins magnitude system (Sirianni et al 2005). Objects classified as being good stellar candidates were retained from the {\sc dolphot} output, with all other objects, characterised as too faint, too sharp, elongated or extended, being discarded.  A further cut was performed to remove all objects containing bad or saturated pixels, and stars which had PSF fits with $\chi>2.5$.

\begin{figure}
  \begin{center}
    \epsfig{file = 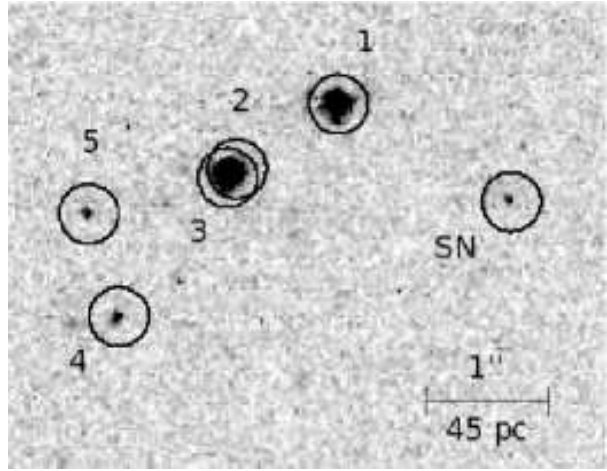, width = 80mm}
    \caption{HST image showing SN~2002ap and several neighbouring stellar objects. This image is the sum of the F435W (Aug 2004) and F814W (Jul 2004) HRC observations. Although objects 2 and 3 appear to be a single source in this image, they are detected as two PSFs separated by 3 ACS/HRC pixels (0\farcs075) by the PSF-fitting photometry package {\sc dolphot}. See text for further details.}
  \label{fig:SNcluster}
  \end{center}
\end{figure}

In the 2004 observations SN~2002ap was detected with a significance of $\sim$ 8$\sigma$ in the F814W image, a significance of $\sim$ 5$\sigma$ in the F555W image, but was not detected in the F435W and F625W images down to their 5$\sigma$ detection limits. Photometry of the SN and of several nearby stars (see Figure~\ref{fig:SNcluster}) is given in Table \ref{tab:dolphot}.
Also shown are 5$\sigma$ detection limits for each of the HST observations. Although objects 2 and 3 in Figure~\ref{fig:SNcluster} appear to be a single object they are detected by {\sc dolphot} as two separate PSFs. Proper scaling of the F435W image allows us to verify that at least two sources exist, where object 2 appears as a much fainter extension to the bright source number 3. From the positions measured by {\sc dolphot} the two sources are separated by $\sim$ 3 ACS/HRC pixels (0\farcs075). The diffraction limited resolution of HST results in stellar PSFs with FWHMs of 0\farcs05 in F435W and 0\farcs09 in F814W. The much larger magnitude difference and the increased blending of objects 2 and 3 as observed in the F814W image makes it impossible to distinguish the two sources by eye in this image. Since Figure~\ref{fig:SNcluster} is a combination of the F435W and F814W observations, objects 2 and 3 appear to be a single source.

Further late-time images of SN~2002ap were taken at the 4.2-m William Herschel Telescope in late August 2006 (see Table \ref{tab:obs} for details). $B$ and $R$ band observations with exposure times matching those of the pre-explosion CFHT images were made using the the Auxiliary Port Imaging Camera (AUX), which has a pixel scale of 0\farcs11/pixel. The reduction and subsequent combining of the images was performed using standard tasks within {\sc iraf}. There were 8$\times$600s exposures taken in the $B$ band on 2006 August 25th. Inspection showed the combined image quality to be 0\farcs75. A set of 6$\times$600s $R$ band exposures were taken on 2006 August 27th, and the quality of the final combined image was 0\farcs7.

\begin{table*}
  \caption[]{Photometry of SN~2002ap and nearby stars in the HST observations of 2004 (see Figure~\ref{fig:SNcluster}). Transformation from HST flight system magnitudes to Johnson-Cousins magnitude system made using the transformation of Sirianni et al. (2005). Also shown are the 5$\sigma$ detection limits derived for each filter at the location of the SN. The detection limits were transformed to standard magnitudes assuming zero colours.}
    \begin{center}
      \begin{tabular}{rcccc} \hline\hline
        & B & V & R & I\\
        \hline
5$\sigma$ limit & 26.3 & 25.7 & 25.4 & 25.6\\
\hline
1 &  25.43(0.10) & 23.48(0.04) & 22.16(0.03) & 21.45(0.01)\\
2 &  24.49(0.06) & 24.60(0.09) & 24.86(0.18) & 24.87(0.17)\\
3 &  22.72(0.02) & 22.06(0.02) & 21.72(0.02) & 21.45(0.01)\\
4 &  24.85(0.07) & 25.29(0.13) & 24.89(0.15) & 25.22(0.16)\\
5 &  25.30(0.09) & 25.56(0.16) & - & -\\
SN & - & 25.84(0.19) & - & 25.05(0.13)\\
        \hline\hline
      \end{tabular}\\
    \end{center}
    {\footnotesize Figures in brackets are the photometric errors}
  \label{tab:dolphot}
\end{table*}

\begin{figure*}
 \centering
  \begin{minipage}[c]{0.5\textwidth}
   \centering
    \epsfig{file = 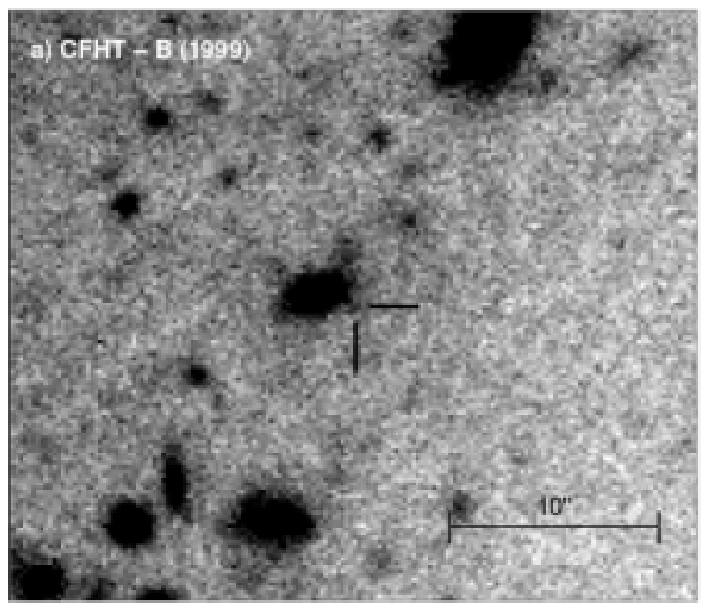, width = 80mm}
  \end{minipage}\\[10pt]
  \begin{minipage}[c]{0.5\textwidth}
    \centering
    \epsfig{file = 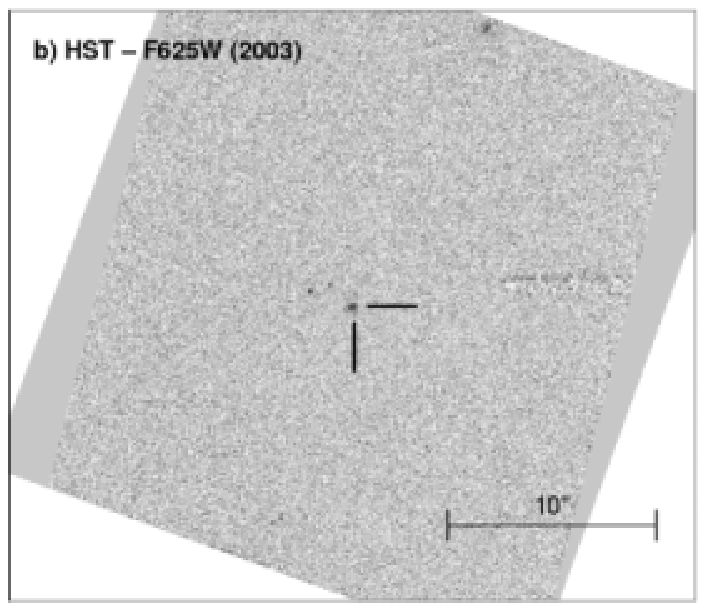, width = 80mm}
  \end{minipage}\\[10pt]
  \begin{minipage}[c]{0.5\textwidth}
   \centering
    \epsfig{file = 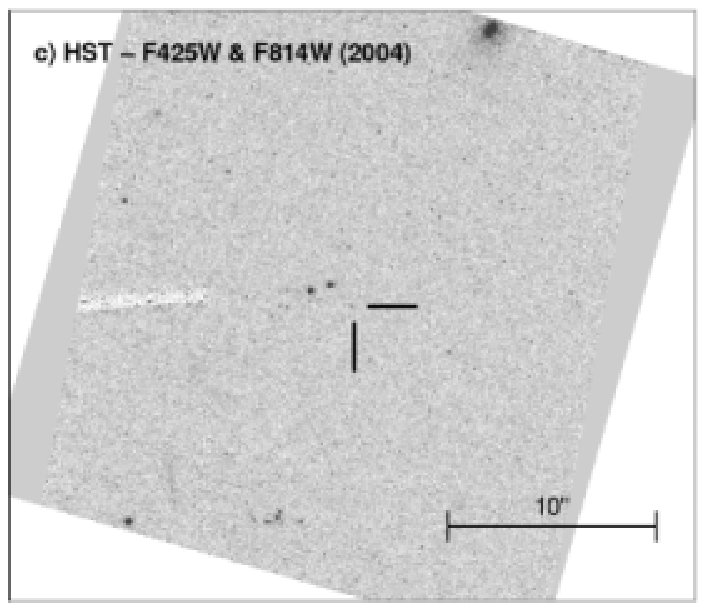, width = 80mm}
  \end{minipage}\\[10pt]
  \caption{Pre- and post-explosion imaging of the site of SN 2002ap.  All images are centred on the SN location. The top panel is the pre-explosion CFHT $R$ band observation, the middle panel is the post-explosion HST F625W observation (120 seconds) taken in Jan 2003, and the lower panel is the sum of the post-explosion HST F814W (720 seconds) and F435W (840 seconds) observations taken in Jul/Aug 2004. Note that the F625W image is much shallower than either of the constituent observations that make up the bottom panel, which explains why it is difficult to identify objects common to both these images. The F625W observation is shown here to convince the reader of the location of SN~2002ap. The SN has faded significantly between the 2003 and 2004 observations, but as can be seen from Figure~\ref{fig:SNcluster} it is still visible at the later epoch.} \label{fig:prepost}
\end{figure*}

\begin{figure*}
  \begin{center}
    \epsfig{file = 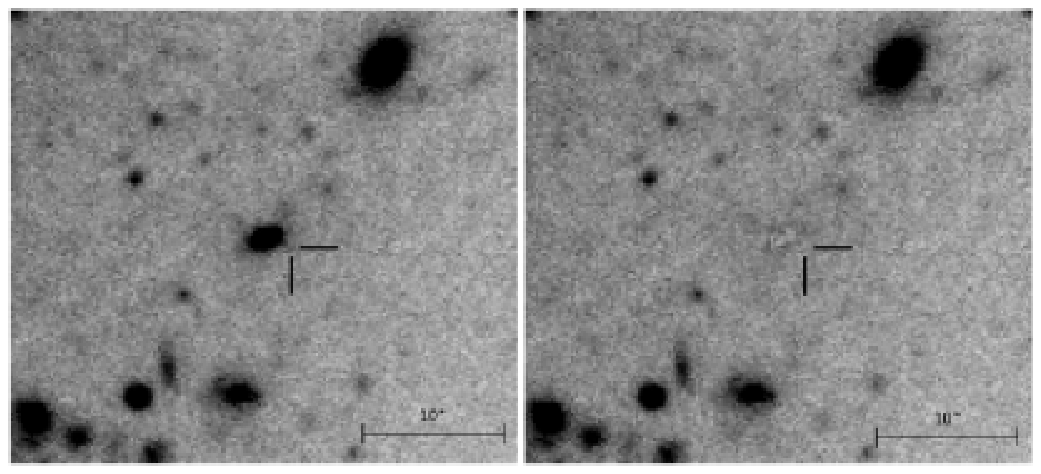, width = 160mm}
    \caption{The original pre-explosion observation (left), and the resultant image after PSF fitting and subtraction of objects close to the SN site (right).}
  \label{fig:subtracted}
  \end{center}
\end{figure*}

\section{Data Analysis}\label{sec:analysis}

\subsection{Alignment of pre- and post-explosion observations}\label{sec:alignment}

The site of SN~2002ap was found to lie on chip 8 of the CHF12K instrument in all the pre-explosion observations. All the images at CFHT were dithered by the observers to improve rejection of hot pixels and other CCD defects. Hence we registered the images in each filter by applying simple linear shifts and then combined them using standard tasks within {\sc iraf}. The stacked $BVR$ images were then co-aligned using the $R$-band image as the reference frame. To determine the precise location of SN~2002ap on the CFHT pre-explosion images, we employed the same method as used by, for example, Smartt et al. (2004) and Maund \& Smartt (2005). 

In order to calculate a transformation to map the ACS/HRC (post-explosion) coordinate system to the CFHT (pre-explosion) observations, the positions of stars common to both sets of images were required. It proved particularly difficult to identify common stars due to the extreme differences in image resolution (CFHT $\sim$0\farcs8; HST $\sim$0\farcs08) and the pixel scales of the two instruments (CFH12K 0\farcs2/pixel; ACS/HRC 0\farcs025/pixel). Most of the stars in the HRC field of view are in resolved clusters, which appear blended in the lower resolution ground based observations. Such stars cannot be used to align the images since reliable measurements of their positions cannot be made in the CFHT frames. Nevertheless ten isolated stellar objects common to both the pre- and post-explosion observations were identified, and their positions measured using the centring algorithms within the {\sc iraf} {\sc daophot} package. 

A general geometric transformation (x and y shifts, scales and rotations) which mapped the ACS/HRC coordinate system to that of the CFH12K instrument was calculated using the {\sc iraf} task {\sc geomap} and the positions of those stars common to both sets of observations. Figure~\ref{fig:prepost} shows sections of the pre- and post-explosion images centred on the SN coordinates and aligned using the {\sc geomap} solution. The pixel position of SN~2002ap in the post-explosion F814W frame was measured using the three centring algorithms of the {\sc daophot} package (centroid, ofilter, gauss), and the mean value transformed to the coordinate system of the CFHT images using the {\sc iraf} task {\sc geoxytran}. The error in the SN position was estimated from the standard deviation of these three measurements to be $\sim$ 0\farcs003. This error proved insignificant when compared to the RMS error of the transformation, which was $\sim$ 0\farcs072.

\subsection{The SN site in pre-explosion observations}

The location of the SN site in the CFHT images was investigated to discover if a progenitor star had been observed. Figure~\ref{fig:prepost}a shows that the SN position lies on the edge of a bright source, which is blended in the CHFT frames but, as can be seen from the HST imagery (Figure~\ref{fig:SNcluster}), is formed from at least five separate objects. The FWHM of stellar sources in the CFHT  $BR$ frames is $\sim$ 0\farcs8 and in $V$ $\sim$ 1\farcs2. The SN was found to lie some 2\arcsec\ from the nearest of the sources that constitute the blend, so any progenitor star in the CFHT images should be resolved, provided the observations were deep enough for it to be detected. 

Visual inspection showed there to be significant flux close to the SN location in the $B$ and $R$ frames, with nothing visible in the much shallower $V$ band image. At first sight this appears to be a detection of a progenitor star. To test this, two methods were used to determine whether or not the progenitor star had been recovered.

\subsubsection{PSF fitting photometry using {\sc DAOPHOT}}\label{sec:psffitting}

Method 1 utilised the PSF photometry tasks within the {\sc iraf} {\sc daophot} package, and followed the techniques for crowded field photometry described by Massey \& Davis (1992). Aperture photometry was initially performed on objects across the CFHT frames, before an empirical model point spread function (PSF) was created from several isolated stars using the task {\sc psf}. Several different PSF models were created in order to test how sensitive the photometry results were to the model used. The task {\sc allstar} was run on a subset of stars in the CFHT frames for each PSF model, and the photometry results from each run compared. The results varied little with the PSF model used, and could be considered identical to within the photometric errors. The positions of the SN and of the nearby stars were measured in the HST images and transformed to the CFHT coordinate system using the transformation discussed in Section~\ref{sec:alignment}. These positions, along with the magnitudes measured by {\sc dolphot}, were fed into the {\sc allstar} task, which performed simultaneous PSF fitting on all the objects, including any possible progenitor at the SN site. The success of the fitting procedure was judged from the fit residuals in the subtracted image. Figure~\ref{fig:subtracted} shows an original CFHT frame and a PSF subtracted version.  

The {\sc allstar} task was implemented in two modes: 1) allowing re-centring of objects during PSF fitting using the initial input coordinates as starting positions, and 2) no re-centring of objects. A summary of the results is presented in Table~\ref{tab:allstarphot}. 

In the case where re-centring was permitted the position of the object recovered closest to that of the SN was significantly shifted from the supernova's precise location; 0\farcs435 in the $B$ band image and 0\farcs368 in the $R$ band. The positions of this object in the pre-explosion observations were measured using several methods and mean values were calculated for each filter. One measurement was made by fitting all of the sources simultaneously. Three more measurements were made by first of all subtracting out all of the sources except the possible progenitor, and then measuring its position using the three centring algorithms of the {\sc iraf} {\sc daophot} package. One final measurement was made by fitting a PSF to the source in this subtracted image. The positional errors were estimated from the standard deviation of these five values and are shown in Table~\ref{tab:astrometric_errors} along with the other astrometric errors discussed in Section~\ref{sec:alignment}. From the total astrometric errors appropriate for each filter we find 4.1$\sigma$ and 2.5$\sigma$ differences between the position of the progenitor site and the object positions measured in the $B$ and the $R$ band images respectively. Moreover, the positions measured in the pre-explosion $B$ and $R$ band frames are themselves separated by $\sim$ 0\farcs200. The astrometric error with which to compare this separation is independent of the transformation error, and is simply a combination of those associated with measuring positions on each of the CFHT frames; 0\farcs148. Therefore 0\farcs200 is equivalent to a separation of $\sim$ 1.4$\sigma$, which is too high to confirm coincidence yet too low to rule it out. Equally the displacements of the pre-explosion sources from the SN position are of such significance that it is impossible to draw any definite conclusions on coincidence.

\begin{table}
  \caption[]{{\sc daophot} PSF fitting photometry of supernova site in pre-explosion CFHT observations.}
    \begin{center}
      \begin{tabular}{rcc} \hline\hline
        Filter & $\Delta$posn (mas) &Mag\\
        \hline
\multicolumn{3}{l}{Recentering during profile fitting}\\[1.5ex]
B & 435 & 25.47(0.10)\\
R & 368 & 24.80(0.14)\\
	\hline
\multicolumn{3}{l}{No recentering during profile fitting}\\[1.5ex]
B & 0 & 25.85(0.19)\\
R & 0 & 25.00(0.19)\\
        \hline\hline
      \end{tabular}\\[1.5ex]
    {\footnotesize Figures in brackets are the photometric errors}
    \end{center}
  \label{tab:allstarphot}
\end{table}

However it is possible that the re-centred positions of these objects are unreliable. It has been found that while allowing stars to be re-centred during profile fitting, the positions of some of the faintest stars may be greatly shifted towards peaks in the subtraction noise of nearby very bright stars (Shearer et al. 1996). Since no objects corresponding to those found in the CFHT images were observed in the HST frames, and since the shift from the SN position was towards the nearby bright stars, it is reasonable to suggest that the above might have happened in this case.

The alternative is therefore to prohibit re-centring, forcing each PSF fit to be centred at the input coordinates throughout the fitting process. Via this method it was possible to fit a PSF at the progenitor site in both the $B$ and the $R$ band images. The magnitudes of these fits were slightly fainter than those from the re-centring case; $\sim$ 0.35 mags fainter in $B$ and $\sim$ 0.20 mags fainter in $R$ (Table~\ref{tab:allstarphot}).  However, by disabling re-centring we lost any independent verification of the object's position, since we forced the PSF fit at the predetermined SN position. It is therefore not possible to conclude from PSF fitting techniques that the progenitor star of SN~2002ap is detected in the pre-explosion images, although the fact that the visible flux cannot be explained by any objects in the HST post-explosion observations makes such a conclusion appealing.

\begin{table}
  \caption[]{Astrometric error associated with the alignment of pre and post-explosion observations.}
    \begin{center}
      \begin{tabular}{lr} \hline\hline
Error & Value (mas)\\
\hline
SN position & 3\\
Geometric transformation (RMS) & 72\\
\\
Pre-explosion position - $B$& 78\\
Pre-explosion position - $R$& 126\\
\hline
Total Error - $B$& 106\\
Total Error - $R$& 145\\ 
        \hline\hline
      \end{tabular}\\
    \end{center}
  \label{tab:astrometric_errors}
\end{table}

\subsubsection{Comparison with ground based late-time observations}\label{sec:imagesubtraction}

\begin{figure*}
  \begin{center}
    \epsfig{file = 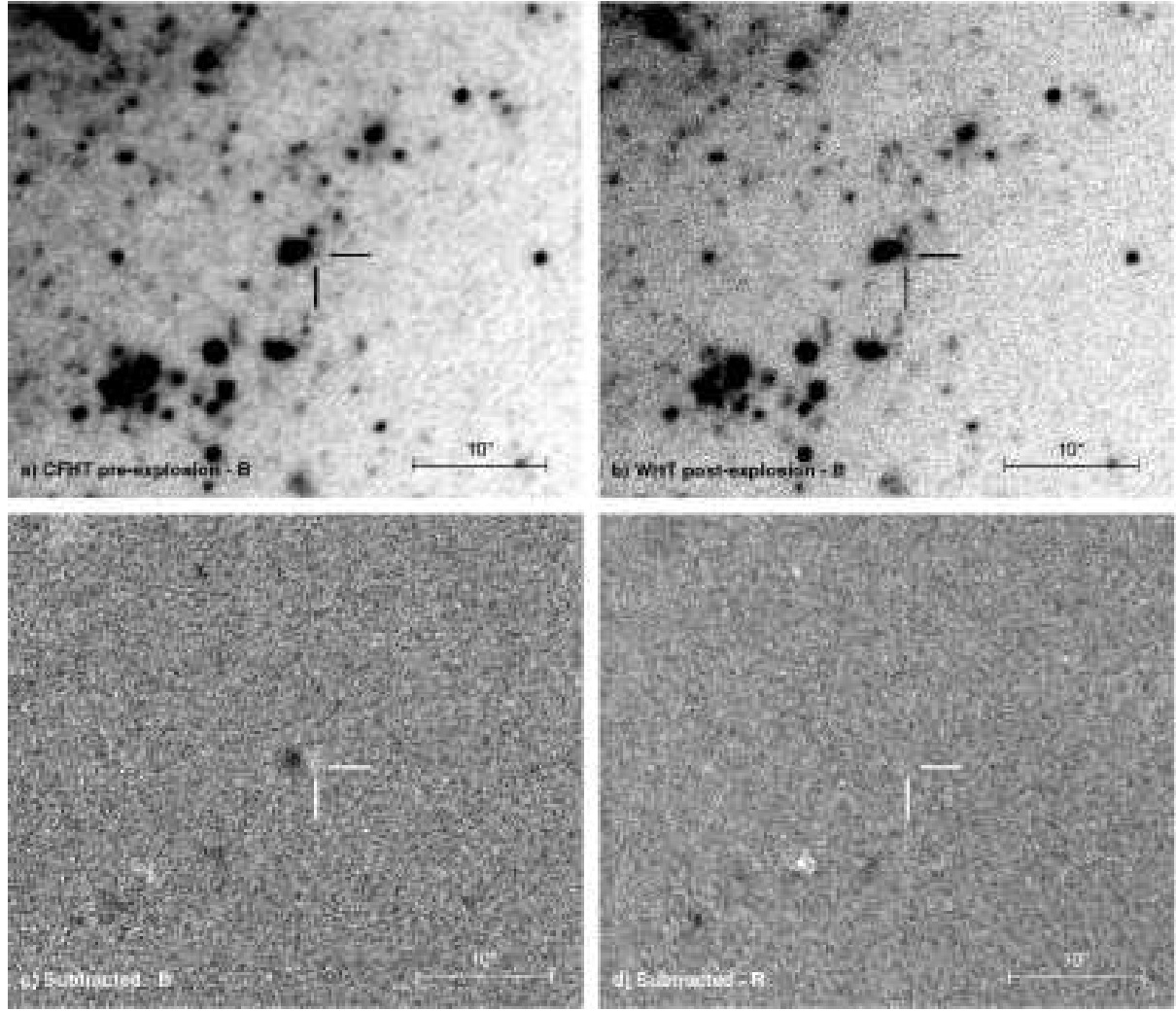, width = 160mm}
    \caption{a) CFHT pre-explosion $B$ observation, b) WHT post-explosion $B$ observation, c) $pre$ $minus$ $post$ subtracted $B$-band image and d) $pre$ $minus$ $post$ subtracted $R$-band image. All the images are shown with inverted intensities (positive sources appear black, negative sources appear white). The $R$-band subtraction (d) appears to be perfectly flat at the position of the SN, which is marked in all frames by cross hairs. In the $B$-band subtracted image (c) a faint and extended negative (white) feature is visible overlapping the SN location. Adjacent to this, coincident with the centre of the blend of nearby stars, is a bright positive (black) source. It is believed that both of these subtraction residuals are due to differences in the transmission functions of the CFHT and WHT filters (see Figure~\ref{fig:filtfunctions}), where the negative feature overlapping the SN position is a diffuse source of emission line flux (possibly a H\,{\sc ii} region) that is more efficiently tranmitted by the WHT $B$ filter, and the positive source is due to extra continuum flux from the nearby bright stars that is transmitted by the CFHT filter.}
  \label{fig:isis_subtraction}
  \end{center}
\end{figure*}

Another method that can be used to detect the presence of a progenitor star in archival imagery is to subtract a late-time post-explosion image from the pre-explosion version. The progenitor star will obviously not exist in the post-SN images, and as long as the SN itself has faded enough to contribute negligible flux, the subtraction of post-SN imagery from pre-SN frames should reveal a positive source at the progenitor position. This of course depends on whether the original observations were deep enough for the progenitor star to be detected above the background noise. To perform such an image subtraction late-time observations of SN~2002ap were taken with the Auxiliary Port Imaging Camera (AUX) of the 4.2-m William Herschel Telescope (WHT) (details in Table \ref{tab:obs}). Observations were made using $B$ and $R$ band filters, with identical exposure times and in similar seeing conditions to the CFHT observations. Taken approximately four and a half years post-explosion, the SN was much too faint to have been detected. (Note the magnitudes and detection limits for the SN in the 2004 HST observations - Table~\ref{tab:dolphot}.) Visual inspection showed the depth of the pre- and post-explosion images to be very similar (see Figures~\ref{fig:isis_subtraction}a and~\ref{fig:isis_subtraction}b). 

After re-binning the WHT frames to match the pixel scale of the CFH12K camera (CFH12K pixel scale 0\farcs2; AUX pixel scale 0\farcs1) and accurate image alignment, the image subtraction package {\sc isis} 2.2 (Alard \& Lupton 1998; Alard 2000) was used to subtract the post-explosion images from their pre-explosion counterparts. Subtraction of the $R$-band observations resulted in a near perfectly flat subtracted image at the position of the SN (see Figure~\ref{fig:isis_subtraction}d), implying that the same source is visible pre- and post-explosion and therefore cannot be the SN progenitor. However, in the case of the $B$-band subtraction an extended negative feature, albeit of quite low signal-to-noise, is visible overlapping the SN position (see Figure~\ref{fig:isis_subtraction}c). Given the sense of the subtraction ({\it pre minus post}) a negative feature indicates an increase in detected flux in the post-explosion $B$-band image, an observation that is inconsistent with the detection of a progenitor in the pre-explosion frame. Also visible in the $B$-band subtracted image is a bright positive residual coincident with the centre of the adjacent blended source some 2\arcsec\ from the SN position, implying that something in this blend appeared brighter in the pre-explosion image. We can instantly rule this out as a progenitor detection based solely on the astrometry. 
One might argue that the progenitor is detected in the pre-explosion frames and that the flux in the WHT post-explosion $B$ and $R$-band observations is from the SN, which has rebrightened perhaps due to interaction with circumstellar material or a light echo from interstellar dust. This scenario could potentially explain the differences between the $B$-band images. On the other hand it is highly unlikely. Comparing the $B$ and $R$ magnitudes of the pre-explosion source (Table~\ref{tab:allstarphot}) with the 5$\sigma$ detection limits of the SN in August 2004 (Table~\ref{tab:dolphot}) we see that by this epoch the SN has already become significantly fainter than the pre-explosion object. In order to reproduce the subtracted images, the SN would have to become significantly brighter between August 2004 and August 2006 when the WHT images were taken. Crucially, its $R$ magnitude at this later epoch would have to exactly match that of the pre-explosion source, with its $B$ magnitude becoming only slightly brighter. Such coincidental matching of the SN magnitudes to those of the pre-SN source is virtually impossible. 

A simpler explanation is that the source close to the SN position in the pre- and post-explosion images is the same object, with the residuals in the subtracted image arising from differences between the CFH12K and the AUX filter functions. Since this object is still visible post-explosion obviously it cannot be the progenitor. So what is it?

Nothing is visible in the 2004 HST observations at the position of this source, which lies between the SN and the closest of the nearby stars in Figure~\ref{fig:SNcluster}. This is in spite of the fact that the detection limits of the HST images are significantly deeper than the measured magnitudes of this object in the CFHT frames. If, however, we assume that the object is a region of extended emission, we can explain this non-detection by the apparent low sensitivity of the ACS/HRC to such sources when compared with the CFHT instrument. Inspection of Figure~\ref{fig:prepost} shows several areas where there is significant flux in the CFHT images which cannot be accounted for by stars visible in the high resolution HST frames. Some traces of this flux are just visible in the HST F814W observation from 2004. Due to the very small pixel scale of the ACS/HRC (0\farcs025/pixel) the light from a region of extended emission is spread over many pixels and is therefore easily dominated by detector read noise if exposure times are not sufficiently long. In contrast the CFH12K pixels, which cover an area on the sky sixty four times that of the HRC pixels, are much more sensitive for imaging such sources. Summing the HST frames from all filters, and rebinning to match the pixel scale of the CFH12K camera, gave some indication of a diffuse region of emission close to the SN position and extending towards the bright stars nearby, but attempts to estimate its magnitude proved futile due to the very low signal-to-noise ratio.

\begin{figure*}
  \begin{center}
    \epsfig{file = 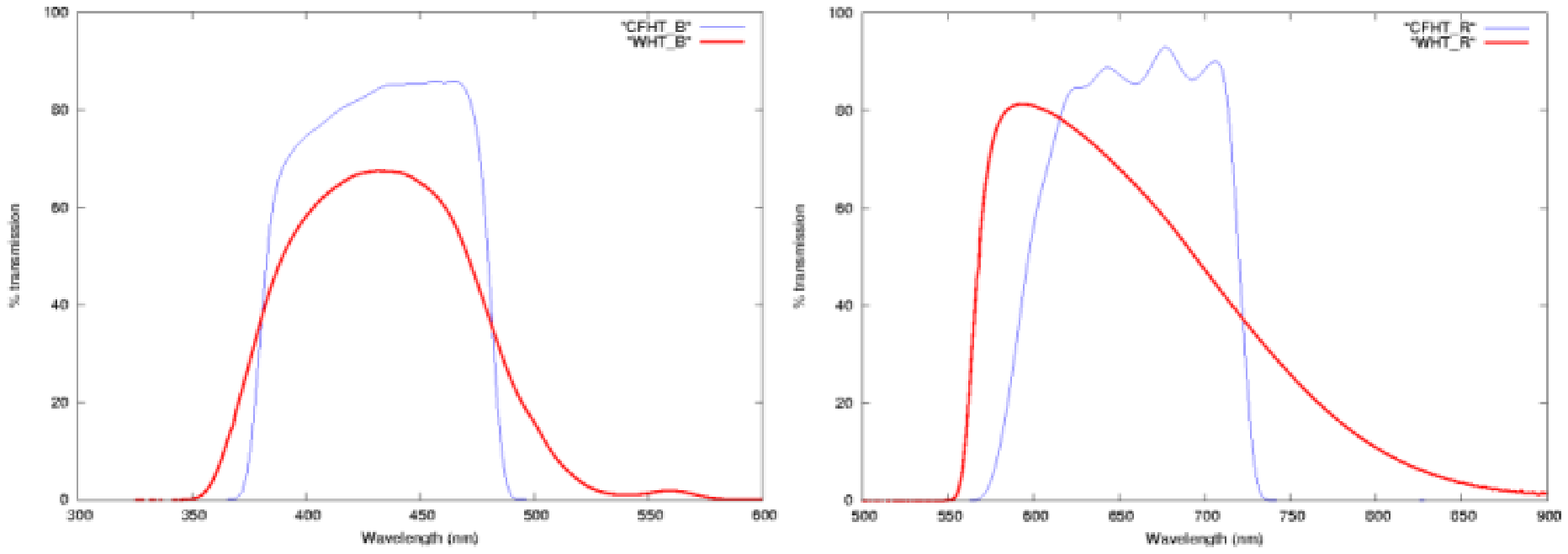, width = 160mm}
    \caption{Comparison of CFHT/CFH12K and WHT/AUX $B$ (left) and $R$ (right) filter functions}
  \label{fig:filtfunctions}
  \end{center}
\end{figure*}

If the flux from this diffuse source is dominated by emission lines (for example a H\,{\sc ii} region) this fact coupled with the differences between the CFHT and the WHT filter functions could explain the negative residuals in the subtracted $B$-band image (see Figure~\ref{fig:filtfunctions}). Unfortunately the $H\alpha$ pre-explosion observation was too shallow (900s compared to the 3600s $R$-band exposure) to produce a significant detection at the SN site, so it cannot be used to confirm the presence of a H\,{\sc ii} region. 

We conclude that the object seen in the pre-explosion observations (in Fig.~\ref{fig:prepost}) close to the SN position is a diffuse source, possibly a H\,{\sc ii} region, and {\it not} the SN progenitor.

\subsection{Detection limits of the pre-explosion observations}\label{sec:limits}

Since the progenitor star was not detected, limiting magnitudes were derived for the CFHT $B$ and $R$ band observations using several methods. 

The standard deviation of the counts per pixel, in an area of 5$\times$5 pixels centred on the SN location, was used as a measure of the noise per pixel at the SN site in each image. By rearranging the equation for Signal-to-Noise ratio

\begin{center}
\begin{equation}
\label{equ:S/N}
S/N = \frac{F_{star}}{\sqrt{F_{star} + \sigma^2}}  
\end{equation}
\end{center}

where $F_{star}$ is the flux of a star and $\sigma$ is the background noise level (both in electron counts), one can calculate what $F_{star}$ must be to produce a detection with a signal-to-noise ratio of a given value. In this case we decided to calculate the $F_{star}$ that would give a 5$\sigma$ detection within a 4 pixel ($\approx$ FWHM of the PSF) radius aperture. An aperture correction was applied to the resultant magnitude to give the final 5$\sigma$ limiting magnitude. The value of $\sigma$ in Equation~\ref{equ:S/N} is the standard deviation in electron counts for the whole 4 pixel aperture, $\sigma_{ap}$, and is calculated by multiplying the measured standard deviation per pixel, $\sigma_{pix}$, by the root of the number of pixels in the aperture, $\sqrt{N}$.

Another method involved measuring $\sigma_{ap}$ directly. A 6$\times$6 grid of 4 pixel radius apertures was used to measure the flux of an area of blank sky adjacent to the SN location. For each aperture the sky background was measured in a concentric annulus and subtracted from the total counts in the aperture. $\sigma_{ap}$ was then calculated as the standard deviation of the remaining counts in the 36 apertures. This method is preferable to using the total counts in each aperture to calculate the standard deviation since total counts can be significantly affected by changes in the level of the smooth background. An aperture at one side of the grid may overlie a region of much higher background flux than an aperture at the other side, producing large count differences and hence a value of $\sigma_{ap}$ that is unrealistically high. Subtraction of the local background at each location produces aperture counts that are more readily comparable, as they are measures of the deviation from an average level. Since we wished to measure the background fluctuations and noise on a scale comparable to the size of the PSF in the images and not large-scale variation in the background it was considered best to use the background subtracted counts.

One final estimate for the 5$\sigma$ detection limits was made using the photometry of stars within a 100\arcsec$\times$ 100\arcsec\ section of each image, roughly centred on the SN position. The 5$\sigma$ detection limit was determined as the average magnitude of all the stars with a photometric error of $\sim$ 0.2 mags.

The detection limits determined from the above methods were averaged to produce 5$\sigma$ limiting magnitudes of $B$ = $26.0\pm0.2$ mags and $R$ = $24.9\pm0.2$ mags. A magnitude limit for the much shallower and less restrictive $V$ band image was not derived.

\section{Distance, extinction and metallicity}\label{sec:dist}

Hendry et al. (2005) used three different methods to estimate the distance to M~74; standard candle method (SCM), brightest supergiants, and a kinematic distance estimate. From the mean of these three methods they estimate the distance to be $9.3 \pm 1.8$ Mpc. A light echo was discovered around another SN which exploded in this galaxy (SN 2003gd) and this study suggests a smaller distance of 7.2 Mpc would be more consistent with a model for the light echo flux \citep{van_dyk_echo}. A further determination of distance from the expanding photosphere method gives a distance of $6.7 \pm 4.5$ Mpc \citep{vinko_epm}. However given the large uncertainty of this result and the fact that the light echo method does not produce a definite independent estimate, we will adopt the distance of $9.3 \pm 1.8$ Mpc, noting that if the distance is lower the luminosity limit of the progenitor would be closer to the lower limit of the uncertainty we derive. 

The total extinction toward SN2002ap has been shown by all previous studies to be low \citep{smartt_02ap, vinko_epm, foley_phot02ap, mazzali_02ap} and we adopt the value of \ebv = 0.09 $\pm$ 0.01 measured from a high resolution spectrum of the SN by Takada-Hidai et al. (2002). Assuming the reddening laws of Cardelli et al. (1989), with $R_V$ = 3.1, gives $A_B$ = 0.37 $\pm$ 0.04 and $A_R$ = 0.21 $\pm$ 0.02.

The metallicity gradient of M74 has been studied from the nebular emission lines of HII regions by Van Zee et al. (1998) and we have estimated the oxygen abundance from the latest emission line strengths combined with the latest empirical calibration from Bresolin et al. (2004). The abundance gradient in M74 suggests that at the galactocentric distance of SN2002ap \citep{smartt_02ap} the metallicity is 12 + log [O/H] = 8.4 $\pm$ 0.1. This latest calibration tends to give lower abundances for significantly metal rich regions than have previously been derived, but does not significantly change that originally derived in Smartt et al. (2002). Another recent study of the metallicity gradient in M74 is that of Pilyugin et al. (2004). At the position of SN~2002ap this would suggest a metallicity of $8.3\pm0.1$ dex. Hence it appears that the environment of SN2002ap is mildly metal deficient, somewhat similar to the massive stars in the Large Magellanic Cloud, which have an oxygen abundance of 8.35 $\pm$ 0.1 dex \citep{hunter_lmc}. However we note that Modjaz et al. (2007) determine a slightly higher value of 8.6 dex using a different calibration of the oxygen line fluxes.

\section{The progenitor of SN~2002ap}

From the detection limits of Section~\ref{sec:limits} and the values of distance and extinction from Section~\ref{sec:dist} we derive 5$\sigma$ absolute magnitude limits for the progenitor of SN~2002ap of \mbox{$M_B$ $\geq$ -4.2 $\pm$0.5} and \mbox{$M_R$ $\geq$ -5.1 $\pm$0.5}. The uncertainties are the errors in the sensitivity limits ($\pm$0.2), distance modulus ($\pm$0.43) and extinction ($A_B \pm$0.04; $A_R \pm$0.02) combined in quadrature. Although no progenitor star is detected, these limits constitute the tightest constraints yet placed on the absolute magnitude of a Type Ic SN progenitor, and as such can be used to deduce the likely properties of the star. The properties of greatest interest are the type of star that exploded and its mass, and several methods were employed to estimate these from the data.

\subsection{Single star evolution models and luminosity limits}\label{sec:single_stars}

\begin{figure*}
  \begin{center}
    \epsfig{file = 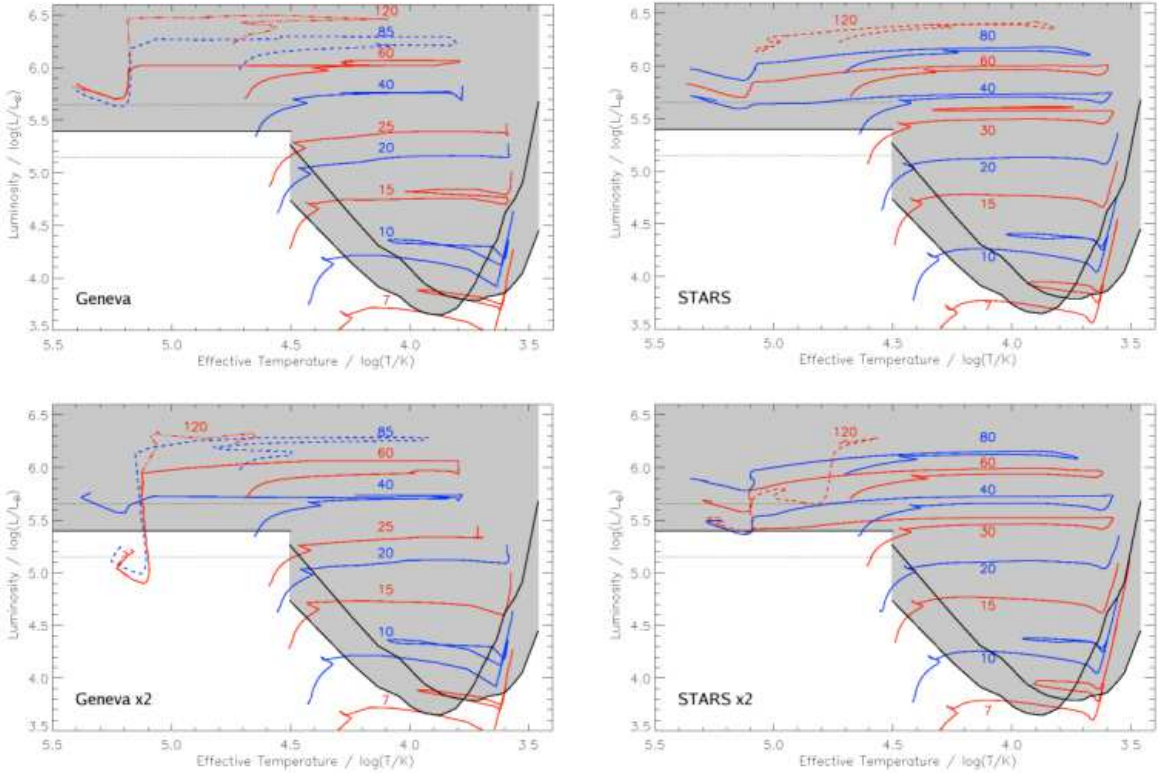, width = 160mm}
    \caption{H-R diagrams showing the 5$\sigma$ luminosity limits for the progenitor of SN~2002ap and stellar evolutionary tracks for singly evolving stars with initial masses of 7-120$M_{\odot}$. Top left and top right show the Geneva and Cambridge STARS stellar models respectively, both for metallicity Z=0.008 (approximately LMC) and incorporating standard mass loss rates. Bottom left and bottom right show the Geneva and STARS models at the same metallicity but with double the standard mass loss rates during the pre-Wolf-Rayet and WNL phases of evolution. Any star within the shaded area on each plot would have been detectable in one or both of the $BR$ pre-explosion observations. Only those models falling outside of this shaded area can be considered as viable progenitors for SN~2002ap.}
  \label{fig:H-R_diag}
  \end{center}
\end{figure*}

Since we had no knowledge of the object's colour, we calculated 5$\sigma$ luminosity limits from $M_B$ and $M_R$ for a range of supergiant spectral types using the colours and bolometric corrections of Drilling \& Landolt (2000). This method is the same as that used by, for example, Maund \& Smartt (2005). The luminosity limits were plotted on Hertzsprung-Russell (H-R) diagrams (Figure~\ref{fig:H-R_diag}), along with stellar evolutionary tracks for single stars of initial mass between 7 and 120$M_\odot$. Four sets of stellar models were used, all of approximately LMC metallicity (metal/mass fraction Z=0.008); two sets from the Geneva group (Schaerer et al. 1993, Meynet et al. 1994) and two sets created using the Cambridge STARS\footnote{http://www.ast.cam.ac.uk/$\sim$stars} code (Eldridge et al. 2006). Of the two sets of models from each code, one incorporated standard mass-loss rates and the other arbitrarily doubled mass-loss rates during the pre-WR and WNL phases of evolution.  Models from both groups were used in order to gauge any model dependence in our interpretation. Any supergiant star belonging to the solid shaded regions above the luminosity limits in these H-R diagrams would have been detected in one or more of the CFHT images, and can therefore be ruled out as a possible progenitor. In this way all evolved hydrogen-rich stars with initial masses greater than 8$M_\odot$ can be ruled out as possible progenitors. Such a conclusion is entirely consistent with the spectral characteristics (lack of H-lines) of SN~2002ap.

At effective temperatures ($T_{\rm eff}$) higher than those of the hottest O-stars are the evolved stages of stars with initial masses greater than $\sim$25-30$M_\odot$. At this late stage of their evolution such stars have lost all or most of their H-rich envelope due to mass loss processes, such as a strong stellar wind, explosive outbursts or interaction with a binary companion. The resulting objects (Wolf-Rayet (W-R) stars) are the exposed helium cores of the original massive stars. W-R stars are divided into two main subtypes; WN stars, in which the emission spectra are dominated by lines of nitrogen and helium, and WC stars, where carbon and oxygen lines dominate. The most oxygen rich WC stars are often further classified as WO. These sub-types are best explained as an evolutionary sequence of the form $WN \rightarrow WC \rightarrow WO$ (Conti 1982). The enhanced helium and nitrogen abundances observed in the spectra of WN stars correspond well to the equilibrium products of H-burning via the CNO cycle, which appear at the stellar surface as a result of mass loss. Subsequent He-burning and further mass loss results in the enhancement of carbon and oxygen at the stellar surface along with the depletion of helium and nitrogen. The star has become a WC star. Still more advanced stages of helium burning result in an overabundance of oxygen at the expense of carbon, producing a WO star.  

SN theory predicts that the progenitors of Type Ib/c SNe are W-R stars, since Ib(Ic) spectra show no signs of H(H or He), which is most easily explained by the progenitor's lack of a H-rich(He-rich) envelope. A theoretical Mass-Luminosity relationship exists for W-R stars \citep{maeder_83, smith_maeder_89, crowther_07}. Crowther (2007) states this relationship for a H-free W-R star as   

\begin{center}
\begin{equation}
\label{equ:mass-lum}
\log \frac{L}{L_\odot} = 3.032 + 2.695 \log \frac{M}{M_\odot} - 0.461 \left( \log \frac{M}{M_\odot} \right)^2.
\end{equation}
\end{center}

Using this formula one can calculate the mass of a W-R star if its luminosity is known. Note this calculation yields the mass of the W-R star, not the initial mass of the star from which it evolved. In this case we can calculate an upper limit for the W-R star's mass from the sensitivity limits of the CFHT observations. However to convert our $M_B$ and $M_R$ limits to luminosity limits requires appropriate bolometric and colour corrections. Since it is not possible to constrain the colours of the progenitor star from magnitude limits we instead consider the broadband colours of W-R stars in the LMC (Massey 2002), the vast majority of which have colours in the ranges $-0.4\leq B-V\leq0.6$ and $-0.1\leq V-R\leq0.4$. Bolometric corrections range from -2.7 to -6.0 \citep{crowther_07}, and Smith \& Maeder (1989) find a BC of $-4.5\pm0.2$ as appropriate for a wide range of galactic W-R stars. The bolometric correction is, however, related to the narrowband $v$ magnitude of the star, as photometry of W-R stars is usually performed with narrowband filters in order to sample the continuum flux between strong emission features. Differences between narrow (denoted by lowercase letters) and broadband photometry of W-R stars can be considerable due to the extra flux contributed by emission lines in the broadband observations. For example, $b-B$ = 0.55 and $v-V$ = 0.75 have been found for the strongest lined WC stars (Smartt et al. 2002). Here we use our broadband sensitivity limits and note that the actual bolometric magnitude limits will be equal to or fainter than what is calculated. For example if $b-B$ = 0.55 is applicable for the progenitor, then the limiting $M_{bol}$ should be fainter by 0.55 mag.

Approximate values of $M_{bol}$ were calculated from the magnitude limits $M_B$ and $M_R$ using a BC of -4.5 and the median of the colours of the LMC W-Rs appropriate in each case. 

$M_B \rightarrow M_{bol} > -8.80\pm0.60$

$M_R \rightarrow M_{bol} > -9.45\pm0.56$

The errors are a combination of the estimated uncertainties in the absolute magnitudes ($\pm0.5$), bolometric correction ($\pm$0.2), and colours ($B-V \pm 0.26$; $V-R \pm 0.17$). The luminosity limit derived from the most constraining value of $M_{bol}$ is $\log (L/L_{\odot}) < 5.40 \pm 0.24$. Using equation~\ref{equ:mass-lum} this limit corresponds to a final W-R progenitor mass of $<12^{+5}_{-3}M_{\odot}$, which is entirely consistent with that of $\sim$ 5$M_{\odot}$ found by Mazzali et al. (2002) from modelling of the SN explosion.

It may be possible to infer a likely range of initial masses through comparison with stellar models as we have done for supergiant stars. For completeness our W-R luminosity limit is plotted in Figure~\ref{fig:H-R_diag} as a straight line across each plot at high temperatures. This is a somewhat crude approximation but since the error is conservative, the upper limit allows for equally conservative constraints to be placed on the star's properties. Considering the models with standard mass loss rates first, the Geneva and the STARS codes predict W-R stars to form only from stars initially more massive than about 34$M_{\odot}$. These independent stellar evolution codes appear to be in good agreement, although there are noticeable differences during W-R evolution phases. These differences are manifested in the initial mass versus final mass functions produced by each code, the final luminosity of each model W-R star being representative of final mass. The STARS models produce higher final W-R masses with increasing initial mass. The Geneva tracks show a similar behaviour, but reach a peak final mass at an initial mass of 60$M_{\odot}$. The Geneva models of initial mass higher than 60$M_{\odot}$ produce W-R stars of substantially lower final mass than similar STARS models. These differences are due to the lower mass loss rates prescribed during W-R evolution in the STARS code, which uses the W-R mass loss rates of Nugis \& Lamers (2000) and includes scaling of W-R mass loss rates with initial metallicity (Eldridge \& Vink 2006). Despite the discrepancies the overall conclusions drawn from both sets of standard mass loss models are the same. Ultimately at no point during W-R evolution do any of the models fall below the detection limits. This implies that no singly evolving star at metallicity Z=0.008 could have been the progenitor of SN~2002ap.

If single-star evolution is to be used to explain the progenitor of SN~2002ap, then we must invoke enhanced mass-loss in the models. As mentioned previously, mass loss can be a result of several phenomena such as stellar winds, eruptive outbursts and binary interaction. The latter we consider in Section~\ref{sec:binary}. For now we consider only singly evolving stars. Eruptive events such as LBV bursts are poorly understood but may account for substantial amounts of mass lost by very massive stars during their lifetimes (Smith \& Owocki 2006). Wind-driven mass-loss is slightly better understood. Three main factors affect this process: luminosity, surface temperature and composition. Higher luminosities in turn drive stronger stellar winds and increased metallicity also leads to higher mass loss rates (e.g. Abbott 1982). This results in the minimum initial mass for a W-R star decreasing with increasing metallicity. The Geneva group provide Z=0.008 models with double the standard mass loss rates during the pre-W-R and WNL phases of evolution (i.e. while each model retains some hydrogen), and we produced similar STARS models for comparison. The increased mass loss rates may be considered as a correction for an underestimation of the rates at Z=0.008, or of the initial metallicity of the progenitor of SN~2002ap. A factor of two increase in mass-loss rates is believed to represent a realistic upper limit, given the intrinsic uncertainty of the rates and the uncertainty in our measurement of metallicity. Comparing the luminosity limit to these tracks provides robust limits on what types of single stars, evolving in the canonical fashion, might reasonably have been the SN progenitor. The Geneva tracks with $\times$2 standard rates suggest the progenitor would have to have been initially more massive than 40$M_{\odot}$. Taking into account the upper error on the luminosity limit, the STARS $\times$2 models suggest that the progenitor could have evolved from a single star with initial mass in the range 30-60$M_{\odot}$, or a star of 120$M_{\odot}$. The differences between the Geneva $\times$2 and STARS $\times$2 models are again due to the different W-R mass-loss rates employed by each code.

It is interesting to compare our analysis here with the work of Maund, Smartt \& Schweizer (2005). In an attempt to place limits on the progenitor of another Type Ic supernova, SN~2004gt, they calculated colour and bolometric corrections for W-R stars by performing synthetic photometry on model W-R spectra from Grafener et al. (2002) (see Maund \& Smartt 2005 for details). In this manner they derived a typical bolometric correction of \mbox{$\sim$ -2.7}. Applying a similar BC to the limiting magnitudes presented in this paper would push our W-R luminosity limit down by 0.7 dex to a median value of $\log(L/L_{\odot})$ = 4.7. The upper W-R mass limit inferred in this case would be around 5$M_{\odot}$, much more restrictive than the 9$M_{\odot}$ limit found by Maund, Smartt \& Schweizer (2005) for the progenitor of SN~2004gt. This revised luminosity limit would rule out all single star progenitors, even from those models with double the standard mass loss rates. Our choice of a BC of -4.5 has been explained above and although the limits produced using this value are conservative in comparison, we believe they offer the most realistic interpretation given the model uncertainties.       

\begin{figure*}
  \begin{center}
    \epsfig{file = 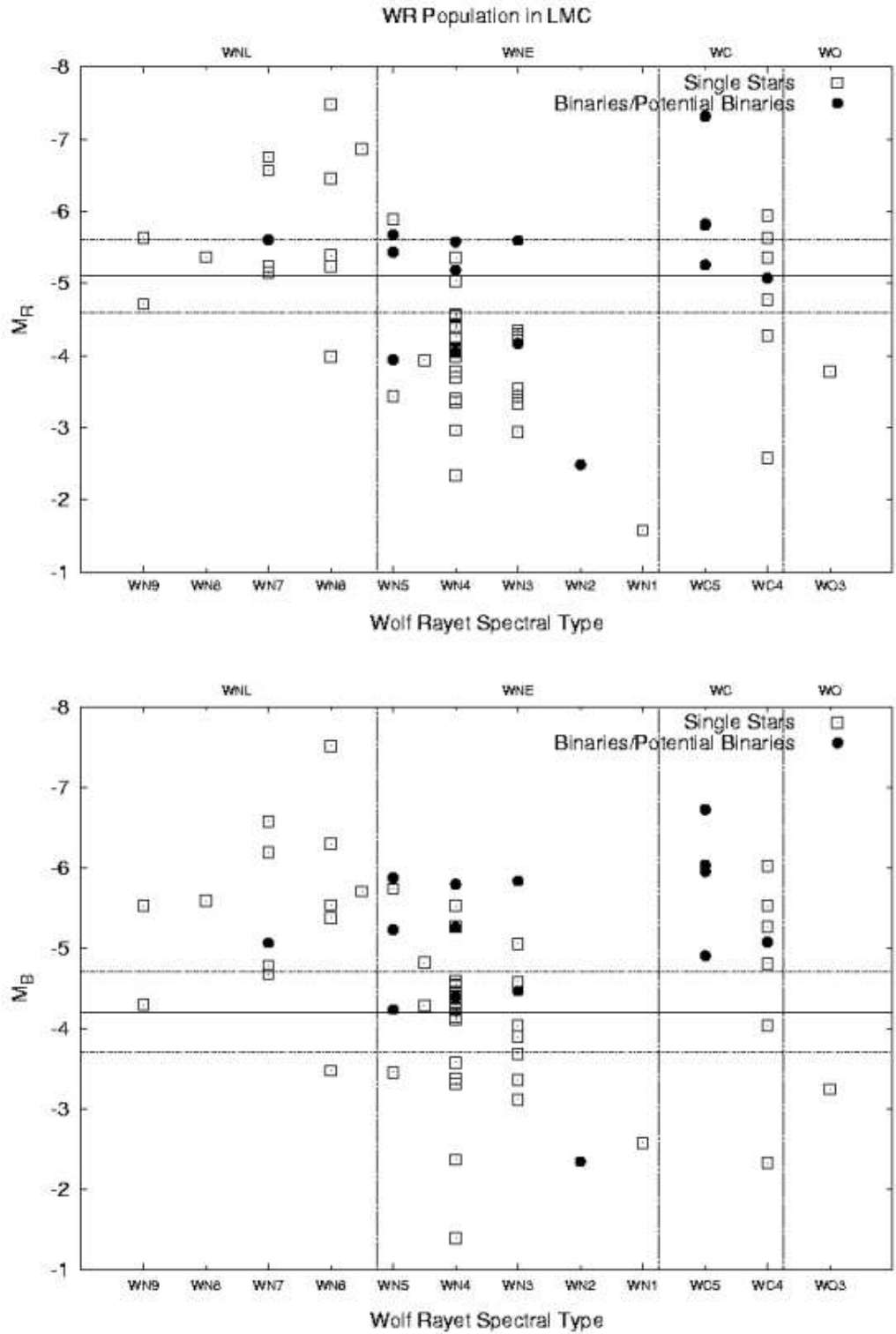, width = 140mm}
    \caption{Broadband photometry of W-R stars in the LMC (Massey 2002) compared to the absolute $B$ (bottom) and $R$ (top) magnitude limits. Note that the evolutionary sequence for W-R stars is WNL $\rightarrow$ WNE $\rightarrow$ WC $\rightarrow$ WO.}
  \label{fig:LMC_WR}
  \end{center}
\end{figure*}

In Figure \ref{fig:LMC_WR} we compare our $M_B$ and $M_R$ limits with broadband photometry of W-R stars in the LMC (Massey 2002), a comparison that is independent of BC. Considering the more restrictive $M_B$ limit, around 79 per cent of WNL stars can be ruled out as possible progenitors along with 23 per cent of WNE stars and 75 per cent of WC and WO stars. Given that SN~2002ap was a Type Ic SN, and therefore deficient of both hydrogen and helium, one would expect the W-R progenitor to have been an evolved WC or WO star. Assuming this to be true we can say that the progenitor was amongst the faintest $\sim$ 25 per cent of WC/WO stars which are known to exist. However our sensitivity limits cannot directly rule out around 21 per cent of the WNL stars and 77 per cent of the WNE stars as possible progenitors. WNL stars, which mark the transition of normal stars to W-R stars, have not yet lost all of their hydrogen envelope and can almost certainly be ruled out as progenitors based on the SN classification. Since such a large fraction of observed WNE stars fall below our detection limits, and with uncertainty as to how much helium must be present to result in a Type Ib instead of a Type Ic SN, we cannot definitely rule out a WNE star progenitor. 

\subsection{Combining observational constraints}\label{sec:indirect_constraints}

Besides trying to constrain the properties of the progenitor of SN~2002ap through direct observation, there are various indirect methods which can be used. One such approach is to observe the SN itself. Through modelling the explosions of carbon-oxygen (C+O) stars Mazzali et al. (2002) created best fits to the lightcurve and spectra of this SN, and thereby estimated that around 2.5$M_{\odot}$ of material was ejected during the explosion. Including the mass of the compact stellar remnant suggests a C+O progenitor star mass of $\sim$5$M_{\odot}$, which would form in a He core of $\sim$7$M_{\odot}$. To produce a helium core of this mass requires a star of initially greater than $\sim$15-20$M_{\odot}$. (This initial mass estimate is lower than the 20-25$M_{\odot}$ suggested by Mazzali et al. (2002) due to the inclusion of convective overshooting in the STARS stellar models.)

X-ray and radio observations of the SN can provide some constraints. The interaction of the fast moving supernova ejecta with the circumstellar material (CSM) leads to emission at X-ray and radio frequencies. Interaction with a higher density of CSM results in stronger emission. However, the radio and X-ray fluxes detected for SN~2002ap soon after explosion were low (Berger et al. 2002; Sutaria et al. 2003; Soria et al. 2004; Bj{\"o}rnsson \& Fransson 2004) suggesting a low density for the material immediately surrounding the star. If we assume that the final mass of the progenitor was around 5$M_{\odot}$, and that it was initially more massive than 15-20$M_{\odot}$, then the star must have lost 10-15$M_{\odot}$ during its lifetime. This material must be sufficiently dispersed prior to the SN explosion so as not to result in strong radio and X-ray emission. Only two epochs of X-ray observations of the SN are available, the first taken less than five days post-explosion and the second about one year later (Soria et al. 2004). Radio observations were taken on a total of 16 epochs, 15 of which spanned the first ten weeks after discovery of the SN (Berger et al. 2002; Sutaria et al. 2003) and one final observation approximately 1.7 years post-explosion in which the SN was not detected (Soderberg et al. 2006). It is reasonable to assume that no interaction occurred between the SN ejecta and any dense CSM during the 18-month gap in the radio monitoring, since such interaction would still have been detectable in the late-time observation (see Fig. 2 of Montes et al. (2000) - the timescale for variation in the radio light curve of SN 1979C is of the order of 4 years). We can say that for at least the first 1.7 years after the explosion of SN~2002ap the ejecta encountered no region of dense CSM. By multiplying this time period by the velocity of the outermost layers of the SN ejecta we can calculate a lower limit for the radius of such CSM. Mazzali et al. (2002) derive a photospheric velocity of $\sim$30,000 \mbox{$\rm{km}\,s^{-1}$} from a spectra taken just 2 days post-explosion (Meikle et al. 2002). However, Mazzali et al. (2002) also suggest that the large degree of line blending in this spectrum requires sufficient material at velocities $>$ 30,000 \mbox{$\rm{km}\,s^{-1}$}, and subsequently employ an ejecta velocity distribution with a cut-off at 65,000 \mbox{$\rm{km}\,s^{-1}$} in their models. Furthermore, through modelling of X-ray and radio observations, Bj{\"o}rnsson \& Fransson (2004) derive an ejecta velocity of $\sim$70,000 \mbox{$\rm{km}\,s^{-1}$}. They also point out that the photospheric velocity only provides a lower limit to the velocity of line-emitting regions (see Section 4 of Bj{\"o}rnsson \& Fransson (2004)). From this range of velocities of 30,000 to 70,000 \mbox{$\rm{km}\,s^{-1}$}, we calculate the minimum radius for a region of dense CSM to be in the range $\sim$ 1.6 to 3.8 $\times$ $10^{12}\,\rm{km}$ ($\sim$ 11,000 to 25,000 astronomical units (au)).

\begin{figure*}
  \begin{center}
    \epsfig{file = 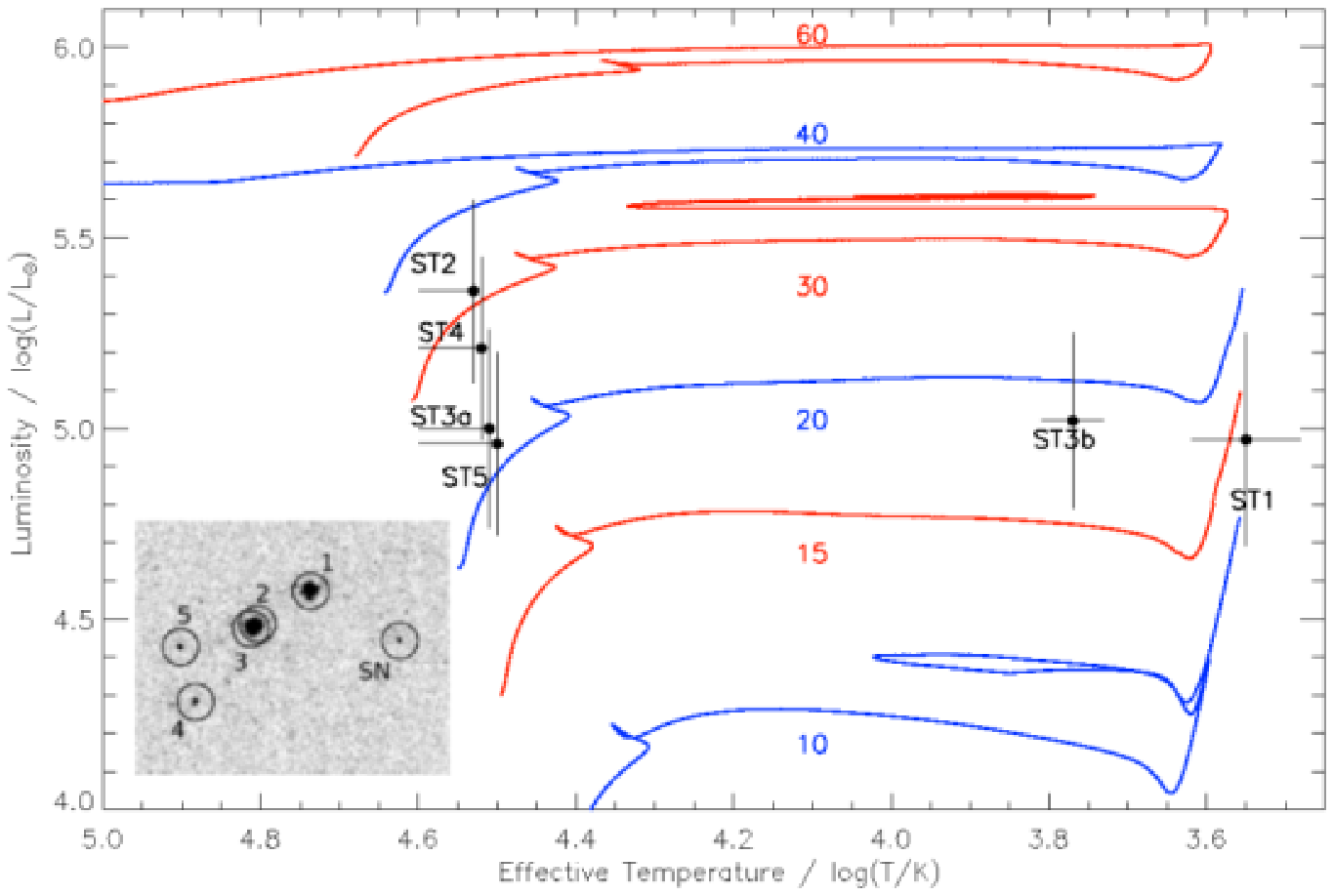, width = 160mm}
    \caption{H-R diagram of stars close to the site of SN~2002ap and STARS stellar evolutionary models of massive stars at metallicity Z=0.008. Since star 3 is {\em not} well fitted by a single SED, this source is better reproduced as the convolution of a double star (3a and 3b). See text for details.}  
  \label{fig:cluster_HR}
  \end{center}
\end{figure*}

One might also attempt to constrain the properties of a SN progenitor by investigating the properties of other stars close to its position. If such stars are found to be coeval with each other and are close enough to the SN site (within a few parsecs) to be considered coeval with the progenitor, then the most evolved of these stars are likely to have had initial masses similar to the SN progenitor. However, in this case SN~2002ap and its closest stellar neighbours are separated from each other by distances of around 50 to 100 pc, as can be seen from the HST observations (see Figure~\ref{fig:SNcluster}). We cannot reliably assume that that these stars are coeval with each other or with the SN progenitor, and therefore cannot use their properties to place any strong constraints on the progenitor star.  

Nevertheless photometry was performed on these stars as detailed in Section~\ref{sec:post} and the results are presented in Table~\ref{tab:dolphot}. Absolute $BVRI$ magnitudes of each star were plotted against wavelength and compared with spectral energy distributions (SEDs) of supergiant stars of spectral type O9 to M5 (Drilling \& Landolt 2000) to determine an approximate $T_{\rm eff}$ and appropriate bolometric correction for each star. The details of how objects 2 and 3 are spatially resolved, at least in the F435W image, have already been discussed fully in Section~\ref{sec:post}. However, while fitting SEDs to each of the objects, it was found that no single SED could be fitted to object 3. This may have been due to increased blending of objects 2 and 3 in observations at longer wavelengths, which in turn would have lead to unreliable photometry, or it could indicate that object 3 is in fact the convolution of several sources. Assuming the latter scenario and the simplest solution therein, i.e. that object 3 is a blend of two stars, we created a program to iterate over all possible combinations of pairs of supergiant SEDs to find the best fit to the data. In this way we tentatively estimated the spectral types and magnitudes of stars 3a and 3b. All the stars were plotted on an H-R diagram and compared to stellar models to estimate their masses, which range from 15 to 40$M_{\odot}$ (see Figure~\ref{fig:cluster_HR}). (Note that this range of masses is not affected if we ignore 3a and 3b.) The lifetime of a 15$M_{\odot}$ star is of the order of 15 Myrs, and for a 40$M_{\odot}$ star around 5 Myr. If all the stars formed at about the same time one would expect to find that the most evolved stars are also the most massive. However the two most evolved stars in Figure~\ref{fig:cluster_HR} appear to be at the lower end of this mass range, at around 15-20$M_{\odot}$. This could suggest that the higher mass stars are unresolved multiple objects rather than single stars, but it is more likely a confirmation that these stars are indeed {\em not} coeval. 


In Section~\ref{sec:single_stars} we found that any single star progenitor would have to be initially more massive than 30-40$M_{\odot}$ and that mass loss rates would have to be double the standard for the measured metallicity.
The final masses of such stars given by the two stellar evolution codes are between 10 and 12$M_{\odot}$. This is significantly larger than the 5$M_{\odot}$ estimate of Mazzali et al. (2002), but consistent with the mass we derived in Section~\ref{sec:single_stars} using the W-R mass-luminosity relationship (not surprising considering both estimates are based on our photometric limits). Enhanced mass-loss rates due to continuum driven outbursts (Smith \& Owocki 2006) might play an important role in the evolution of any progenitor star that is initially more massive than 40$M_{\odot}$ and could help to produce lower mass W-R stars. 

Another possibility is a progenitor produced by the homogeneous evolution (Yoon \& Langer 2005) of a 12-20$M_{\odot}$ star. If the rotation of a main sequence star is fast enough, the rotationally induced mixing occurs on a much shorter timescale than nuclear burning. Fresh hydrogen is continually mixed into the core until the entire hydrogen content of the star is converted into helium. In this way the main sequence star transitions smoothly to become a helium star of the same mass. While Yoon \& Langer (2005) do not consider LMC metallicity objects if we extrapolate their results it is reasonable to expect that the most rapidly rotating stars with initial masses in the range 12-20 $M_{\odot}$ will lead to a Type Ic progenitor in the mass range implied by our luminosity limit and the results of Mazzali et al. (2002). Detailed models of rapidly rotating massive stars would be required to test this hypothesis.

\subsection{Binary evolution}\label{sec:binary}

Binary evolution provides a possible alternative to the single star models discussed previously. Such a scenario is suggested by Mazzali et al. (2002) for the progenitor of this SN, where a star of initially 15-20$M_{\odot}$ loses much of its mass through interaction with a binary companion, to become a C+O star of around 5$M_{\odot}$. 
 
\begin{figure}
  \begin{center}
    \epsfig{file = 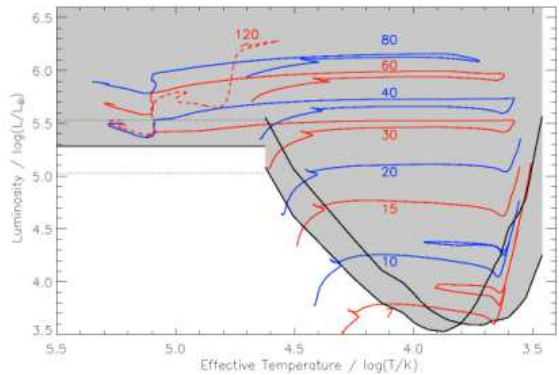, width = 80mm}
    \caption{H-R diagram showing the 5$\sigma$ luminosity limits for a binary companion to the progenitor of SN~2002ap. These limits were derived from the post explosion HST F425W and F814W observations. Also plotted are STARS stellar evolutionary tracks for singly evolving stars with initial masses of 7-120$M_{\odot}$, at metallicity Z=0.008 and with double the standard mass loss rates. Any star within the shaded area would have been detectable in one or both of the aforementioned HST observations. If the progenitor star did have a binary companion it must have been either a main sequence star of $\lesssim 20M_{\odot}$ or a dormant neutron star/black hole.} 
    \label{fig:binary}
  \end{center}
\end{figure}

Using our luminosity limits, it is possible to constrain the binary systems that might have produced the progenitor of SN~2002ap. We can place limits on the luminosity and the mass of the binary companion star in the same way as we did for the progenitor. In the case of the companion, however, we can use both the pre- and the post-explosion observations for this purpose. A companion star was not seen in either set of images, but the post-explosion observations provided slightly more restrictive luminosity limits, and these are plotted on an H-R diagram in Figure~\ref{fig:binary}. Whereas with the SN progenitor we could assume that the star must have been in an evolved state just prior to exploding, we cannot make the same assumption about any possible binary companion. It is conceivable that such a star could be at any stage in its evolution. Comparing our luminosity limits with stellar models we can at least set discrete mass limits for the companion star at various evolutionary phases. 

Considering the case where the binary companion is initially less massive than the SN progenitor, and therefore less evolved at the time of the explosion, we see from Figure~\ref{fig:binary} that it must have been a main sequence star of $\lesssim 20M_{\odot}$. Stars more massive than this would have been visible at all stages of their evolution. Note that this constraint is on the mass and evolutionary phase of the companion at the time of the SN; that is after it has accreted any material from the progenitor. In the case of a star initially more massive than the progenitor, it would have exploded as a SN prior to SN~2002ap leaving a neutron star or black hole as the binary companion\footnote{Nomoto et al. (1994) suggest that an initially more massive companion star may also become a white dwarf in some cases.}. This would of course require the binary to remain bound after the first SN. Brandt \& Podsiadlowski (1995) predict that just 27 per cent of binaries consisting of two high mass stars will remain bound after the initial SN explosion, and of these bound systems 26 per cent will immediately experience strong dynamical mass transfer leading to their merger. Therefore around 20 per cent of high mass binaries might be expected to result in a bound system with a neutron star or black hole component, with only a fraction configured in such a way as to result in the binary interaction required to produce the progenitor of SN~2002ap. Such systems are therefore much rarer than those with main sequence companions.

Using the conclusions of Mazzali et al. (2002) as best estimates of the initial and final masses of the progenitor, we can say that it lost 10-15$M_{\odot}$ of material before exploding. A substantial fraction of this material must have been ejected from the binary system rather than accreting onto a main sequence companion, otherwise it could have gained enough mass to become detectable in the observations (e.g. Maund et al. 2004). We can therefore conclude that the binary mass transfer process must have been non-conservative in cases where the companion was a main sequence star. Furthermore, this non-conservative mass transfer must have occurred early enough in the progenitor star's lifetime to allow the ejected material sufficient time to disperse to a large radial distance. This condition must be met to remain consistent with the X-ray and radio observations of the SN discussed in Section~\ref{sec:indirect_constraints}. For a black hole companion there is no such requirement that mass transfer must be non-conservative since a more massive black hole is no more visible than a lower mass counterpart. The rate at which a black hole can accrete material is however limited (Narayan \& Quataert 2005) and in many cases the process will therefore be non-conservative.   

Kippenhahn \& Weigert (1967) classified binary systems which undergo mass transfer into three types, Cases A, B and C. We shall first discuss each case assuming a main sequence companion of $\leq 20M_{\odot}$. Case A mass transfer is where the donor star begins to transfer mass while still on the main sequence. This occurs in very close binary systems with periods of the order of a day. However, the expansion of the star on the main sequence, which causes it to fill its Roche lobe and dictates the rate of mass transfer, occurs on the nuclear timescale. The companion star has no problem adjusting to accept the material at such a low rate, with the result that Case A mass transfer is in general a conservative process. Our previous argument that conservative mass transfer would result in a main sequence companion becoming visible in our observations therefore suggests that Case A mass transfer is not a viable option. Furthermore, while the donor star is the more massive, conservative mass transfer will result in the shrinking of the binary orbit. Since Case A systems are already initially very close, this will often lead to the merger of the two objects. Where the merger is of two main sequence stars the result will be a single star of mass roughly equal to the sum of its components. This object will then continue its evolution as a single star, for which we have already discussed progenitor scenarios in Section~\ref{sec:single_stars}.   

Case C mass transfer is where the donor begins to transfer mass at the end of core helium burning when it expands rapidly, ascending the giant branch for the second time. Case C therefore occurs in wider binaries with orbital periods of around 100 days or more, ensuring that the donor must expand to become a supergiant before filling its Roche lobe. Mass loss will occur at the thermal or possibly even the dynamical timescale of the donor star as it rapidly expands. The rate at which a companion can accrete this material will be dictated by its thermal timescale. A main sequence companion of lower mass will have a long thermal timescale compared to the donor star, and therefore will only accrete material at a fraction of the rate at which it is being lost by the donor. Consequently much of the donor's mass will be ejected from the binary system so that, in the described model at least, Case C mass transfer will be non-conservative. The mass lost from the system will still be relatively close by when the SN explodes since Case C mass transfer occurs late in the evolution of the donor star. Exactly what distance this material has dispersed to depends on the velocity $v$ with which it is ejected from the binary and the elapsed time $t$ between the end of mass transfer and the explosion of the SN. This distance is given by $D$ $\approx$ 3.15 $\times 10^{9}$($v/100\,\rm{km}\,s^{-1}$)($t/1\,\rm{yr})\,\,{km}$ (or $D$ $\approx$ 21($v/100\,\rm{km}\,s^{-1}$)($t/1\,\rm{yr})\,\,{au}$). The velocity of the ejected material $\rm{v}$ will be similar to the orbital velocity of the accreting star, of the order of 10 - 100 $\rm{km}\,s^{-1}$. In Section~\ref{sec:indirect_constraints} we used the radio observations of the SN to calculate the minimum radial distance out to which no significant density of CSM was encountered by the SN ejecta: $D_{\rm{min}}$ = 11,000 to 25,000 $\rm{au}$. Assuming $v$ = 100 $\rm{km}\,s^{-1}$ requires that mass transfer ended at least $t_{\rm{min}}$ = 500 to 1200 yrs prior to the SN\footnote{The model assumed here is rather simple. In reality the strong stellar wind of the newly formed W-R star would collide with the previously ejected material causing it to accelerate (Eldridge et al. 2006 and references therein). This would have the effect of {\it reducing} $t_{\rm{min}}$ above.}. Since this time period is rather short we cannot rule out all Case C mass transfer scenarios. Binaries with orbital radii approaching the lower limit of what constitutes a Case C system might begin to interact early enough so as to satisfy the above requirement. The low extinction measured towards the SN (see Section~\ref{sec:dist}) probably suggests that the 10 to 15$M_{\odot}$ of material lost prior to explosion had dispersed to much larger distances than the lower limit we estimate from the radio and X-ray data. This would tend to rule out all Case C binaries. One can still, however, invoke a non-spherical geometry (e.g. the rings seen around SN~1987A) that would allow this material to be much closer without causing extinction of the SN provided the line-of-sight to the observer is not obscured. 

Case B mass transfer is where the donor star begins to transfer mass when it first becomes a giant at the end of core hydrogen burning, but before the ignition of central helium burning. Orbital periods of systems undergoing this type of interaction are intermediate to those of Cases A and C. As with Case C, the mass loss will occur on the relatively short thermal or dynamical timescales, making it difficult for a main sequence companion to accrete all this material. Case B mass transfer will happen early enough in the evolution of the donor star to allow sufficient time for the ejected material to disperse prior to the SN explosion. This earlier stripping of its hydrogen envelope will also afford the donor star more time to shed its helium envelope through strong stellar winds and become the low mass WC star predicted to be the progenitor of this SN. Based on this argument we suggest that the mass transfer process in our assumed binary system is more likely to have been Case B than Case C. We conclude that in all possible binary systems for the progenitor of SN~2002ap where the companion star was initially less massive than the progenitor, the companion must have been initially $\leq 20M_{\odot}$ and non-conservative Case B mass transfer is most likely to have occurred.

\begin{figure*}
  \begin{center}
    \epsfig{file = 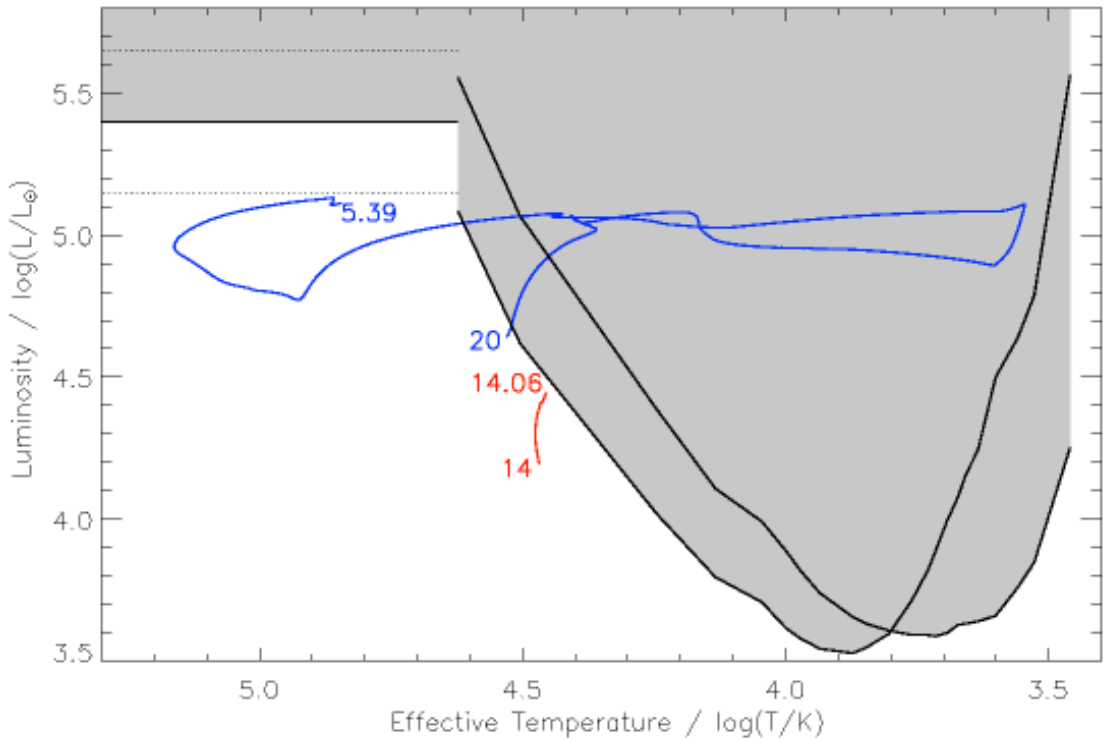, width = 160mm}
    \caption{Binary model of a 20$M_{\odot}$ primary and a 14$M_{\odot}$ secondary created using the Cambridge STARS stellar evolution code and the binary-evolution algorithm of Hurley et al. (2002b). The luminosity limits plotted are derived from the pre-explosion observations in the W-R domain, and from post-explosion observations for O to M-type supergiants. The initial separation is such that the primary fills its Roche lobe as it expands to become a giant after core hydrogen burning, and so Case B mass transfer ensues. However, mass loss through Roche lobe overflow does not remove mass quickly enough to halt the expansion of the primary star, with the result that the primary star expands to engulf its companion. The system has entered a common envelope evolution (CEE) phase. Orbital energy is transferred to the common envelope through an unknown physical process, leading to a reduction in the binary period and separation, and to the ejection of the common envelope. The evolution of both models is halted at the end of core carbon burning in the primary, essentially just prior to the SN explosion. The numbers displayed beside the primary and secondary tracks show initial and final masses. The primary ends its life as a 5.4$M_{\odot}$ WC star. The secondary accretes very little of the $\sim$15$M_{\odot}$ of material lost by the primary, increasing in mass by just 0.06$M_{\odot}$.}  
  \label{fig:binary2}
  \end{center}
\end{figure*}

To test this idea we created binary models using the STARS stellar evolution code (Eldridge, Izzard \& Tout in prep). We note that the modelling of binary stars is extremely uncertain (e.g. Hurley et al. 2002b). In all the models the initial orbital periods were such that Case B mass transfer would occur. One of the models, incorporating a 20$M_{\odot}$ primary and a 14$M_{\odot}$ secondary, is shown as an example in Figure~\ref{fig:binary2} compared to the relevant luminosity limits. The evolution of both models continues up to the end of core carbon burning in the primary. At this point, the time until the primary goes SN is so short that both stars appear essentially as they will when it explodes. Both stars must fall below the luminosity limits at the time of the SN explosion in order to satisfy observations. Being of lower mass, the secondary (the companion) is still on the main sequence as the primary expands and begins to transfer mass. However in this model, and in fact in all of the Case B binary models created, it was found that mass loss through Roche lobe overflow could not remove mass quickly enough to halt the expansion of the primary star, with the result that it expanded to engulf its companion. Such a scenario is referred to as common envelope evolution (CEE). In this example CEE begins with the dense core of the primary and the main sequence secondary orbiting about the centre of mass of the system within the inflated envelope of the primary. Orbital energy is transferred to the common envelope through an unknown physical process, leading to a reduction in the orbital period and radius, and to the ejection of the common envelope. The timescale for CEE is short and therefore very little mass is accreted by the secondary during this time. Provided the binary orbit does not shrink so much as to cause a merger\footnote{Nomoto et al. (2001) propose that some binary mergers of this sort may produce rapidly rotating C+O stars, which could explode as Type Ic hypernovae.}, having been stripped of its H-rich envelope the primary star continues to lose mass through strong stellar winds, eventually shedding most of its He-rich layer also. The model shown in Figure~\ref{fig:binary2} produces a 5.4$M_{\odot}$ C+O W-R from the initially 20$M_{\odot}$ star. 

In the case of an assumed neutron star or black hole companion we derive broadly the same conclusions about mass transfer processes as we do for a main sequence secondary. There are however some notable differences. Case A mass transfer would again most likely lead to the merger of the binary components. However the merger of a main sequence star with a neutron star or black hole would either completely disrupt the progenitor, or possibly result in the formation of a Thorne-Zytkow object (TZO) (Thorne \& Zytkow 1977). Fryer \& Woosley (1998) suggest that the merger of a black hole or neutron star with the He core of a red giant could produce a long complex GRB with no associated supernova. This may occur in Case B systems with initially small orbital separations. In general both Case B and Case C mass transfer processes are possible, either with or without a subsequent CEE phase, although we again suggest that Case B is more likely since it results in the earlier removal of the progenitor's hydrogen envelope and a much longer period of W-R evolution prior to explosion. We are of course unable to place mass limits on a neutron star or black hole companion using our luminosity limits.

Note that in all of the above we have employed binary interaction to remove only the hyrogen envelope of the progenitor star. The subsequent removal of the He-rich layer has occurred via strong radiatively driven winds. Nomoto et al. (2001) point out that under the right conditions a second mass transfer event may occur, which removes the He layer. This however requires a large degree of fine-tuning of the binary system. During the first CEE phase the binary orbit must be reduced enough to allow interaction between the newly formed He-star and its companion, but not so much as to cause the two to merge. This second mass transfer is more likely to occur for lower-mass He-stars since they attain larger stellar radii (Habets 1986). Nomoto et al. (1994) exclusively use such double mass transfer to model the evolution of the progenitor of the Type Ic SN~1994I; a 2.1$M_{\odot}$ C+O star formed from a 4$M_{\odot}$ He-star. The significantly larger estimates for the final progenitor mass of SN~2002ap and the He-core mass from which it formed (see Section~\ref{sec:indirect_constraints}) makes the double mass transfer scheme less likely in this case.

\section{Conclusions}
  
We have used a unique combination of deep, high quality CFHT pre-explosion images of the site of SN~2002ap and follow up HST and WHT images of the SN itself to place the most restrictive limits to date on the luminosity and mass of the progenitor of a Type Ic SN. Theory predicts that the progenitors of such SNe are W-R stars which have lost all of their H and most of their He-rich envelopes. The archival observations rule out as viable progenitors all evolved hydrogen-rich stars with initial masses greater than 7-8$M_{\odot}$, the lower mass limit for stars to undergo core collapse. This is entirely consistent with the observed absence of hydrogen in the spectra of this SN and with the subsequent prediction that its progenitor was a W-R star. The magnitude limits do not allow us to distinguish between WN, WC and WO stars, as examples of all these W-R types have been observed with magnitudes that fall below our detection limits. Since the SN was deficient of helium it is likely to have been a WC or WO star. Using the W-R Mass-Luminosity relationship we calculated the upper limit for the final mass of a W-R progenitor to be $< 12^{+5}_{-3}M_{\odot}$. Comparing the W-R luminosity limit with Geneva and STARS stellar models for single stars at metallicity Z=0.008 we found no viable single star progenitors when standard mass loss rates were considered, and only initially very massive ($>30-40M_{\odot}$) candidate progenitors when mass loss rates were doubled. The lack of significant extinction towards the SN, along with weak detections of the SN at X-ray and radio wavelengths, suggests a relatively low mass of circumstellar material surrounding the progenitor star. This would tend to rule out very high mass progenitors. We conclude that any single star progenitor must have experienced at least twice the standard mass-loss rates, been initially $>30M_{\odot}$ and exploded as a W-R star of mass 10-12$M_{\odot}$.

The most likely alternative to the single star models is a binary evolution scenario, in which the progenitor star is stripped of most of its mass by a companion, becoming a low mass, low luminosity W-R star prior to exploding as a supernova. This idea is consistent with Mazzali et al. (2002) who, based on modelling of the supernova lightcurve and spectra, suggest a binary scenario for the progenitor of SN~2002ap in which a star of initial mass 20-25$M_{\odot}$ becomes a 5$M_{\odot}$ C+O star prior to explosion. (Note that we revise this initial mass to the lower value of 15-20$M_{\odot}$ in Section~\ref{sec:indirect_constraints}.) The luminosity limits from both the pre- and post-explosion observations, along with other observational constraints, allowed us to place limits on the properties of any possible binary companion, and to infer which mass transfer processes could or could not have occurred. We conclude that, if the system was indeed an interacting binary, the companion star was either a main sequence star of $\leq20M_{\odot}$, or a black hole or neutron star formed when an initially more massive companion exploded as a supernova. In either case, mass transfer most likely began as non-conservative Case B, which led immediately to common envelope evolution (CEE).

\section*{Acknowledgments}

This work, conducted as part of the award "Understanding the lives of massive stars from birth to supernovae" made under the European Heads of Research Councils and European Science Foundation EURYI (European Young Investigator) Awards scheme, was supported by funds from the Participating Organisations of EURYI and the EC Sixth Framework Programme. SJS and RMC thank the Leverhulme Trust and the European Science Foundation for a Philip Leverhulme Prize and postgraduate funding. The William Herschel Telescope is operated on the island of La Palma by the Isaac Newton Group in the Spanish Observatorio del Roque de los Muchachos of the Instituto de Astrofisica de Canarias. We thank the Service Programme and Pierre Leisy for rapid coordination of the observations. This research used the facilities of the Canadian Astronomy Data Centre operated by the National Research Council of Canada with the support of the Canadian Space Agency and the Canada-France-Hawaii Telescope (CFHT) which is operated by the National Research Council of Canada, the Institut National des Sciences de l'Univers of the Centre National de la Recherche Scientifique of France, and the University of Hawaii. Based in part on observations made with the NASA/ESA Hubble Space Telescope, obtained from the data archive at the Space Telescope Science Institute. STScI is operated by the Association of Universities for Research in Astronomy, Inc. under NASA contract NAS 5-26555

\label{lastpage}
\end{document}